\documentclass[usenatbib]{mnras}
\pdfoutput=1
\usepackage{times,txfonts}
\usepackage{mathptmx}
\usepackage[T1]{fontenc}
\usepackage{ae,aecompl}

\usepackage{ gensymb }
\usepackage{ wasysym }
\usepackage{float}
\usepackage{lscape}
\usepackage{graphicx}	% Including figure files
\usepackage{amssymb}	% Extra maths symbols

\title[Variable CO ro-vibrational lines in HD163296]{Variability in the CO ro-vibrational lines from HD163296
\thanks{Based on observations made with ESO Telescopes at the La Silla Paranal Observatory under programme ID 088.C-0898A.}}

\author[R. P. Hein Bertelsen et al.]{
Rosina P. Hein Bertelsen$^{1}$\thanks{bertelsen@astro.rug.nl},
I. Kamp$^{1}$,
G. van der Plas$^{2}$,
M.\,E. van den Ancker$^{3}$,
\newauthor L.\,B.\,F.\,M.\,Waters$^{4,5}$,
W.-F. Thi$^{6}$,
P. Woitke$^{7}$
\\
$^{1}$Kapteyn Astronomical Institute, Rijks-universiteit Groningen (RuG), Landleven 12, Groningen 9747, Netherlands\\
$^{2}$Departamento de Astronom\'{\i}a, Universidad de Chile, Casilla 36-D, Santiago, Chile\\
$^{3}$European Southern Observatory, Karl-Schwarzschild-Str.2, D 85748 Garching bei M\"unchen, Germany\\
$^{4}$Anton Pannekoek Astronomical Institute, University of Amsterdam, PO Box 94249, 1090 GE Amsterdam, The Netherlands \\
$^{5}$SRON Netherlands Institute for Space Research, Sorbonnelaan 2, 3584 CA Utrecht, The Netherlands\\
$^{6}$Max-Planck-Institut f\"{u}r extraterrestrische Physik, Giessenbachstrasse 1, 85748 Garching, Germany\\
$^{7}$SUPA, School of Physics \& Astronomy, University of St. Andrews, North Haugh, St. Andrews KY16 9SS, UK\\
}

\date{Accepted to MNRAS January 2016}

\pubyear{2015}

\begin{document}
%\label{firstpage}
%\pagerange{\pageref{firstpage}--\pageref{lastpage}}
\maketitle

% Abstract of the paper
\begin{abstract}
{We present for the first time a direct comparison of multi-epoch (2001--2002 and 2012) CO ro-vibrational emission lines from HD\,163296. We find that both the line shapes and the FWHM (Full Width Half Maximum) differ between these two epochs. The FWHM of the median observed line profiles are 10--25\,km/s larger in the earlier epoch, and confirmed double peaks are only present in high $J$ lines from 2001--2002. The line wings of individual transitions are similar in the two epochs making an additional central component in the later epoch a likely explanation for the single peaks and the lower FWHM. Variations in NIR brightness have been reported and could be linked to the observed variations. 
Additionally, we use the thermo chemical disc code ProDiMo to compare for the first time the line shapes, peak separations, FWHM, and line fluxes, to those observed. The ProDiMo model reproduces the peak separations, and low and mid $J$ line fluxes well. The FWHM however, are over predicted and high $J$ line fluxes are under predicted. 
We propose that a variable non-Keplerian component of the CO ro-vibrational emission, such as a disc wind or an episodic accretion funnel, is causing the difference between the two data sets collected at different epochs, and between model and observations. Additional CO ro-vibrational line detections (with CRIRES/VLT or NIRSPEC/Keck) or [Ne\,{\sc{ii}}] line observations with VISIR/VLT could help to clarify the cause of the variability.}
\end{abstract}

% Select between one and six entries from the list of approved keywords.
% Don't make up new ones.
\begin{keywords}
protoplanetary discs -- line: profiles -- stars: variables: T Tauri, Herbig Ae/Be -- infrared: ISM -- \textit{(stars:)} circumstellar matter
\end{keywords}

%%%%%%%%%%%%%%%%%%%%%%%%%%%%%%%%%%%%%%%%%%%%%%%%%%

%%%%%%%%%%%%%%%%% BODY OF PAPER %%%%%%%%%%%%%%%%%%
\section{Introduction} \label{sec_intro}
Herbig Ae/Be (HAeBe) stars are intermediate mass pre main sequence stars that are surrounded by discs. CO ro-vibrational emission is frequently observed from these discs \citep{blake2004, brittain2007, brittain2009, plas2014} and traces the inner regions close to the star, where planet formation is expected to occur.

HD\,163296 is a well studied Herbig Ae star of spectral type A3Ve \citep{gray1998} located at a distance of 118.6\,pc (Hipparcos). It has been labeled a group II source (flat disc) by \citet{meeus2001} and the stellar mass of the system has been estimated to M$_*$$\sim$2.3\,M$_{\astrosun}$ \citep{montesinos2009}. Using spatially resolved sub-mm data, the inclination of the disc has been estimated to 46$\degree\pm$4$\degree$ and the position angle to 128$\degree\pm$4$\degree$ \citep{Isella2007}. The disc has been studied through interferometric observations of both continuum and CO emission lines (0.87--7.0\,mm) \citep{Isella2007}. From the line emission the authors found the $^{12}${CO} and $^{13}${CO} to be optically thick and derive an outer radius of 550$\pm$50 au. The continuum dust emission implies an outer radius of 200$\pm$15\,au. Thus, a sharp drop in continuum emission of a factor $>$30 at a radius of 200\,au, was invoked to explain the lack of continuum emission from radii larger than 200 au \citep{Isella2007}. {Meanwhile, from a non-detection of the red-shifted jet beyond 500 au radius \citep{grady2000} and the extent of the scattered light emission \citep{Wisniewski2008} the radius of the dust disc can be estimated to 500 au.}

\citet{Tilling2012} presented Herschel/PACS observations of the disc, and used the line detections from this dataset together with additional line and continuum data to fit a ProDiMo  \citep{woitke2009} model, in an attempt to determine disc properties. A continuous disk with inner and outer radii of 0.45 and 700\,au provided a reasonable fit to the optical to millimetre continuum spectral energy distribution, the low an mid $J$ CO line intensities, and the spectral profiles from ground-based interferometric observations.
{ALMA observations (band 7, $\sim$850\,$\mu$m) have put the outer dust disc radius at 240 au and the CO outer radius at 575 au \citep{gregorio2013,rosenfeld2013}. A well resolved dusty disc was seen with no indications of gaps or holes beyond 25\,au.}
\citet{gregorio2013} conclude that a standard tapered-edge model with one unique density profile cannot match both datasets at once. The CO channel maps require a thicker gas disc, while the mid- and far-infrared SED require a flatter dust disc than presented in \citet{Tilling2012}.

CO ro-vibrational v=1--0 emission lines from this disc have been studied on several occasions \citep[][Bertelsen et al. submitted, from hereon B15]{blake2004, salyk2011, brittain2007}. In \citet{salyk2011}, using NIRSPEC data from the Keck telescope, the authors identified double peaked profiles and FWHM$\sim$83\,km/s (for $J$>25) indicating that the origin of the emission is close to the star. Assuming Keplerian rotation, the peak separation measured from the high $J$ profiles would suggest an outer radius for the emission of 2--3\,au.
Recently, we have collected spectra containing CO ro-vibrational emission lines with CRIRES at the VLT, and found single peaked profiles for low $J$ lines while high $J$ lines displayed asymmetric flat topped profiles (B15).

In coronagraphic images a bipolar jet (HH 409) has been seen \citep{grady2000}, with radial velocities for the gas in the jet of 200--300\,km/s.
HD\,163296 has shown variations in NIR (Near Infra Red) brightness (over 30\% in $J$-band, 20\% at $H$-band, and 15\% at $K$-band, de Winter et al. 2001) on timescales of years \citep{sitko2008}. This was seen as an outburst in the 1--5\,$\mu$m wavelength region during 2002. This wavelength region corresponds to a bump frequently seen in SEDs of many HAeBes and is thought to be related to the inner wall of the dust disc \citep{natta2001,dominik2003}. {Variability in this region could be related to an increase of the surface area of the inner wall \citep[puffed up inner wall,][]{sitko2008}}. \citet{sitko2008} found no SED model (based on data from one epoch) that could explain the full nature of this source.

\citet{ellerbroek2014} matched transient optical fading with the enhanced NIR excess \citep[also noticed by][]{sitko2008} and relate these to the jet (HH 409). The authors suggest a scenario where dust clouds, that are launched above the disc plane to heights where they cross the observers line of sight to the star, are responsible for the optical and NIR variability. 
\citet{ellerbroek2014} derive a period of 16.0$\pm$0.7\,years for the increased outflow activity from the jet. Due to epoch-sparsity of the photometric data, it could not be confirmed whether the jet period matches the NIR and optical variations, and thereby that a single mechanism is in fact responsible for both jet and photometric variability.
\citet{klaassen2013} reported the detection of a rotating molecular disc wind that extends to more than 10", seen from ALMA observations of the CO $J$=2--1 and $J$=3--2 lines. 

Using high spectral resolution ($R\sim$12\,000) VLTI-AMBER observations \citet{garcialopez2015} studied the Br$\gamma$ line of HD\,163296 and attempted to constrain  the physical origin of this line. They compared the observations to predictions from a line radiative transfer disc wind model and found that the observations and the modelling both suggest that the Br$\gamma$ emission in this disc comes from a disc wind with launching radii from $\sim$0.02\,au to $\sim$0.04\,au, while the entire disc wind emitting region stretches out to $\sim$0.16\,au.

In this paper we will study the variability in the CO ro-vibrational lines over a period of ten years
(2001--2002 to 2012) and investigate the connection with variability observed in the optical and NIR continuum.
Firstly, we present the three sets of observational data, and the line profiles and FWHM derived from them in a homogeneous way (Sect. \ref{sec:obs} and \ref{sec:obsres}). Hereafter, we compare with modelled CO ro-vibrational lines produced using a previously published disc structure \citep{Tilling2012}, in order to understand the observed CO ro-vibrational line emission (Sect. \ref{sec:mod}). This also tests the predicting power of the model, keeping in mind that it has been fitted to a wide variety of observational data, but never before to the CO ro-vibrational lines themselves. We finish the paper with a discussion of how the line variability can be interpreted in the context of both model and observations (Sect. \ref{sec:disc}) and present the final conclusions in Sect. \ref{sec:conc}.

\section {Observational data} \label{sec:obs}
We use high resolution spectra (R$\sim$100,000) of HD\,163296 collected in March 2012 with the VLT cryogenic high-resolution infrared echelle spectrograph \citep[CRIRES][]{kaeufl2004}. Frome hereon we refer to this dataset as C12. These observations cover a wavelength range from 4.5\,$\mu$m to 5\,$\mu$m with 6 different grating settings (4.6575, 4.7363, 4.9948, 4.6376, 4.8219, and 5.0087\,$\mu$m). 
The details of the data reduction and flux calibration can be found in B15. From C12, we include lines from $J$=0 up to $J$=36. 

For comparison we use high resolution spectra (R$\sim$25000) of HD\,163296 collected with NIRSPEC \citep{mclean1998} on the Keck II telescope. These observations are collected during the period 2001--2002. The spectra from different epochs were combined, resulting in one single spectrum per wavelength region. {This stacking of the lines is necessary in order to improve the S/N on the line profiles. Furthermore, there is no sign of variability between epochs beyond the noise in the spectra (Salyk private communication), hence the combining of the spectra is justified.}
Frome hereon we refer to this dataset as N01/02. Due to low transmission (60--80\%) certain regions were left out of this final spectrum, causing several transitions to have gaps in the centre of their line profiles (see middle frame of Fig. \ref{fig:aalines}).
The observations span a wavelength range from $\sim$4.65--5.15\,$\mu$m and include lines from $J$=1 up to $J$=37. Details on data acquisition and reduction for N01/02 can be found in \citet{salyk2011, salyk2009} and \citet{blake2004}.

We also included additional spectra from NIRSPEC on the Keck II telescope, obtained in March 2002. Frome hereon we refer to this dataset as N02b. The observations span a wavelength region from $\sim$4.64--5.02\,$\mu$m and include lines from $J$=1 up to $J$=33. However, this dataset has several lines distorted by telluric over-correction, due to variable telluric water lines during the night of observations. The data reduction for N02b can be found in \citet{brittain2007}.

For C12, we found a continuum flux at 4.770\,$\mu$m of 16.3 Jy (B15). 
For typical CRIRES spectra, we can expect the accuracy of the flux calibration to be around $\sim$30\% \citep{brown2013}, due to differences in width of the PSF (Point Spread Function) for the science versus the telluric standard spectra (these differences are expected to be due to variations in the performance of the AO system). 
From the N01/02 spectrum a continuum flux at 4.770\,$\mu$m of 12.1 Jy was found. An error of 10--20\% is expected. The continuum fluxes measured for C12 and N01/02 are consistent within the uncertainties.
For BN02b, we do not have flux calibrated spectra.

\section{Observational results and analysis} \label{sec:obsres}
\begin{figure*}
\centering
\begin{minipage}[r]{.3\textwidth}
  \includegraphics[width=.5\textwidth]{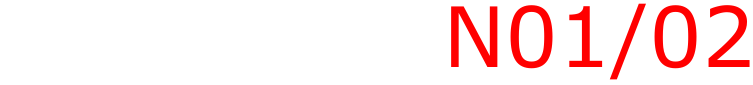}
    \par\vspace{-18pt}
\end{minipage}
\begin{minipage}[r]{.3\textwidth}
  \includegraphics[width=.47\textwidth]{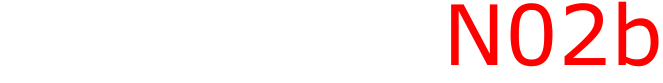}
    \par\vspace{-18pt}
\end{minipage}
\begin{minipage}[r]{.3\textwidth}
  \includegraphics[width=.4\textwidth]{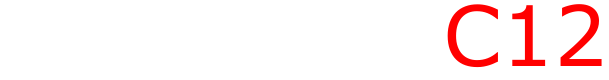}
    \par\vspace{-18pt}
\end{minipage}\\
\begin{minipage}[l]{.3\textwidth}
  \includegraphics[width=\textwidth]{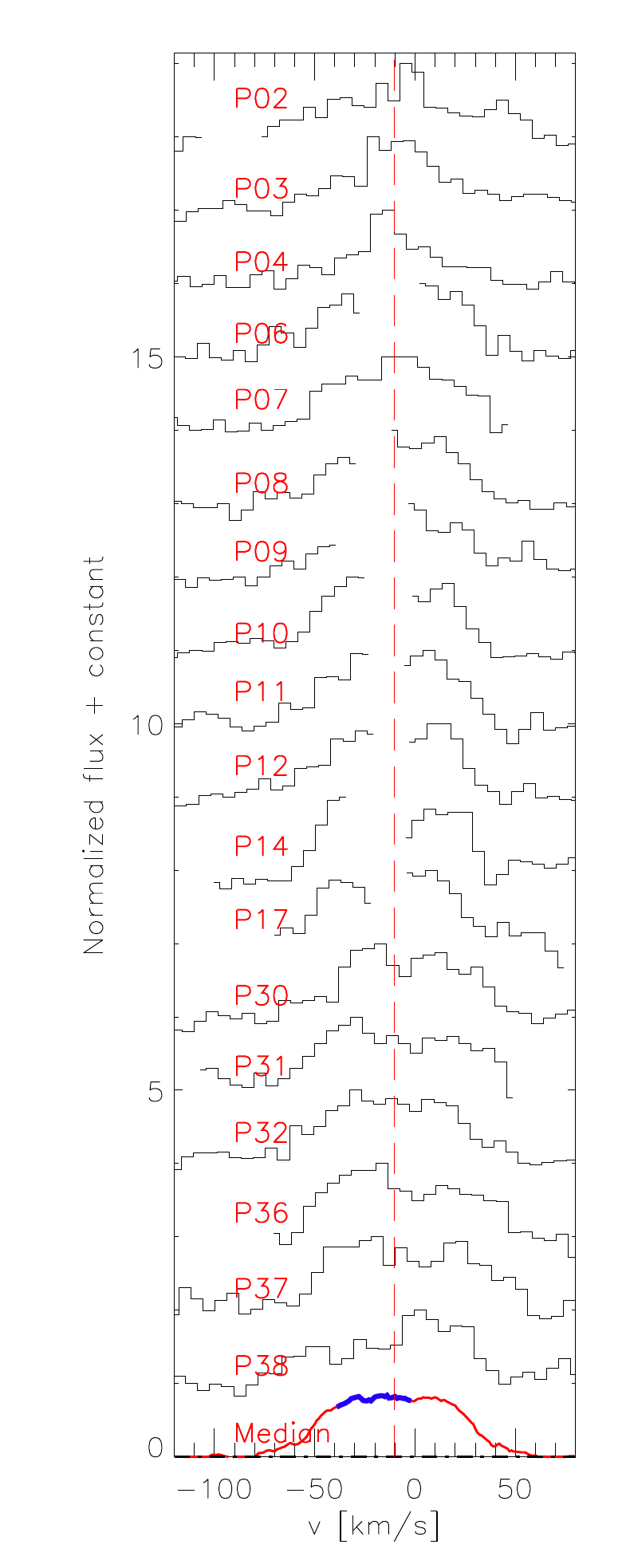}
\end{minipage}
\begin{minipage}[r]{.3\textwidth}
  \includegraphics[width=\textwidth]{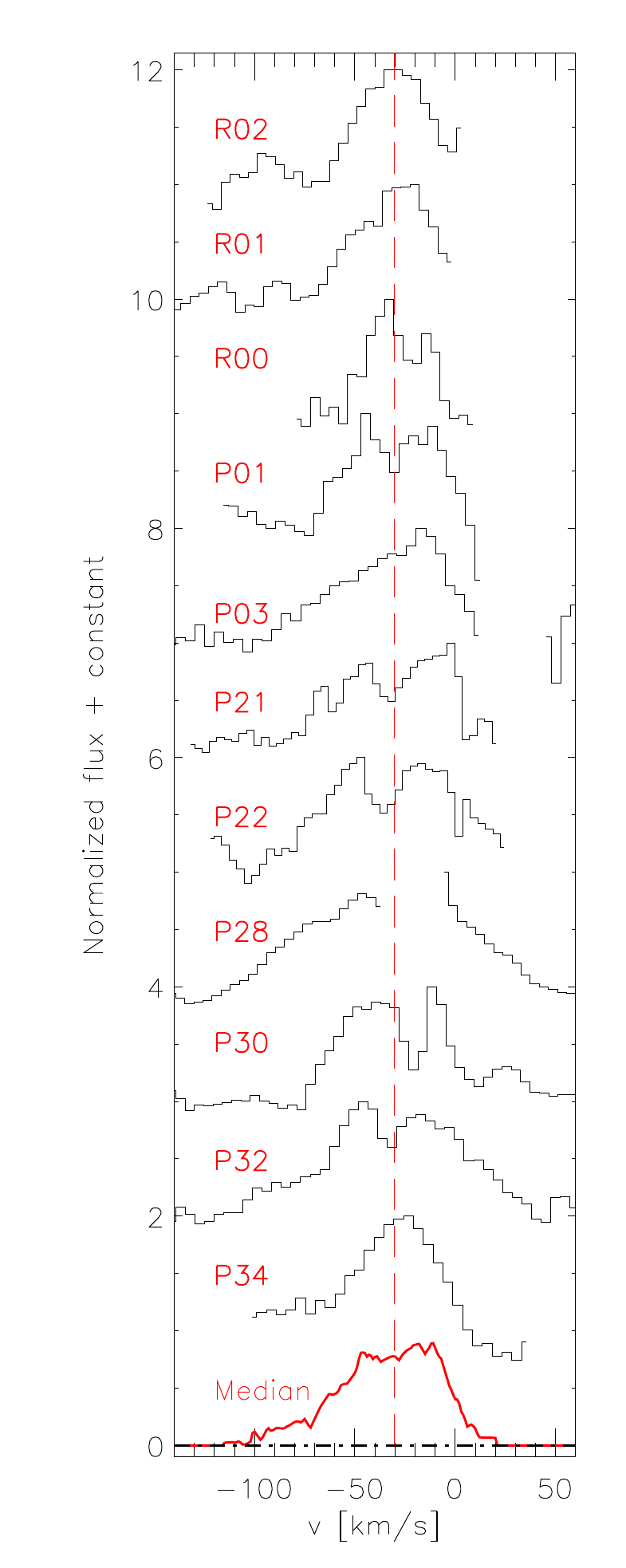}
\end{minipage}
\begin{minipage}[r]{.3\textwidth}
  \includegraphics[width=\textwidth]{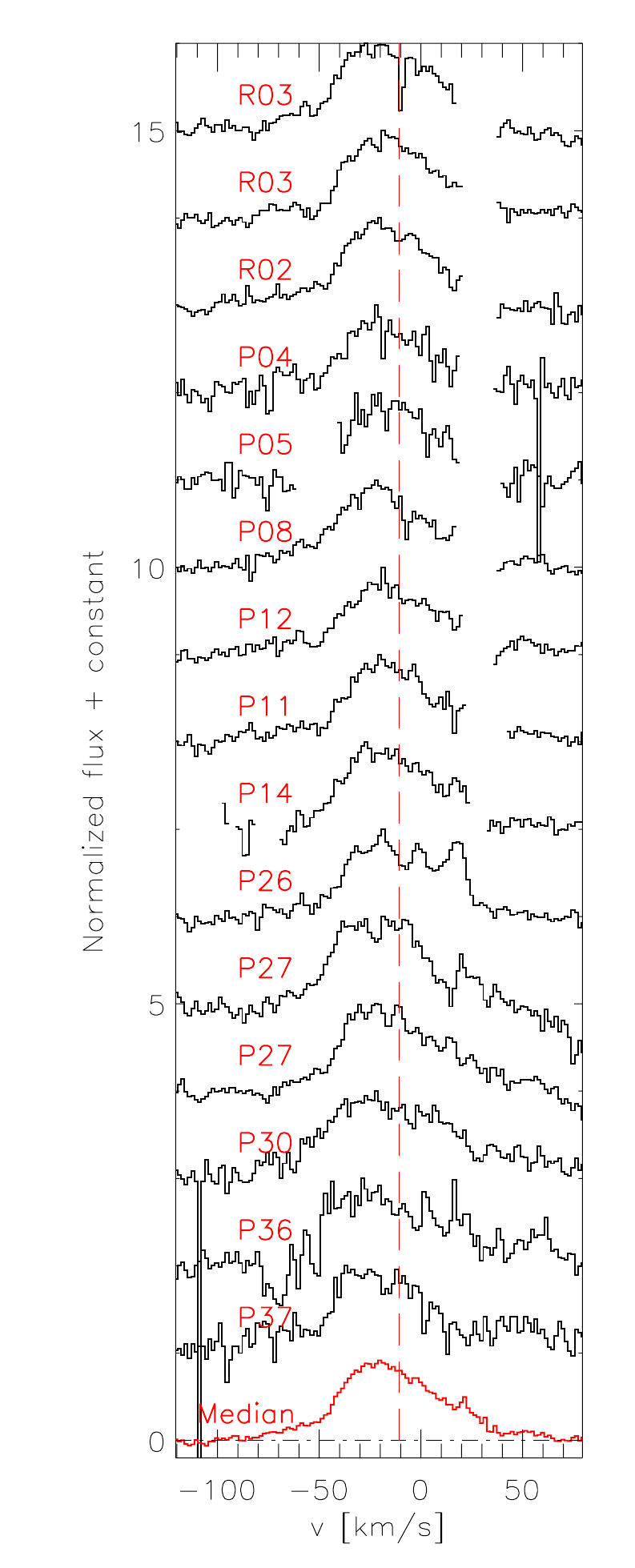}
\end{minipage}
\caption{Individual line profiles and the median at the bottom from all three datasets. \textit{Left}: Line profiles from N01/02. \textit{Middle}: Line profiles from N02b. \textit{Right}: Line profiles from C12. Line identifications are indicated on the plots. Several lines from the N01/02 dataset have 'gaps' at the line centres, thus the region marked in blue on the N01/02 median, corresponding to these missing regions, should be deemed less reliable. Some lines in the C12 dataset are detected twice due to overlapping wavelength settings. }
         \label{fig:aalines}

\centering
\begin{minipage}[l]{.36\textwidth}
  \includegraphics[width=\textwidth]{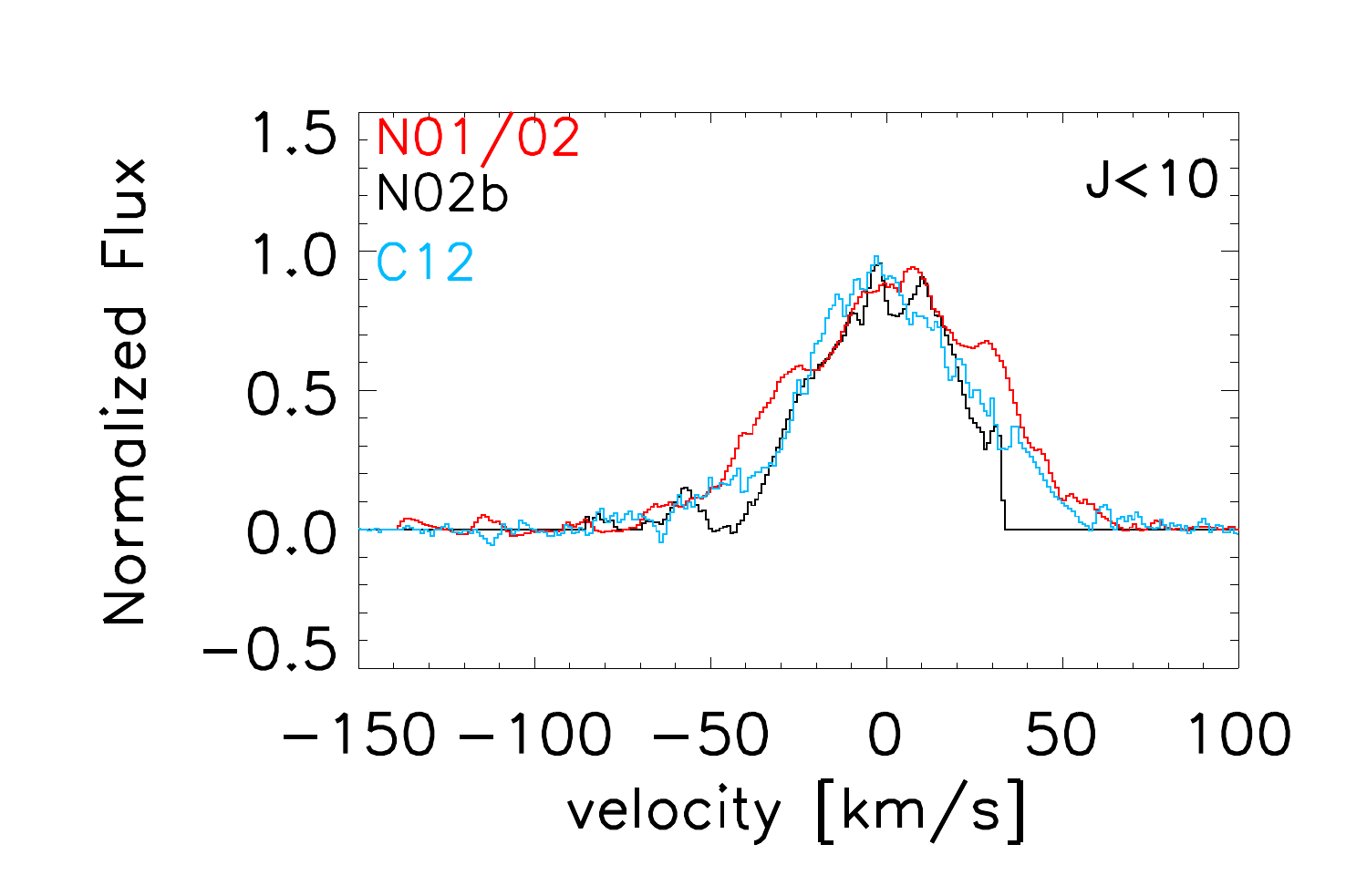}
\end{minipage}
\begin{minipage}[r]{.36\textwidth}
  \includegraphics[width=\textwidth]{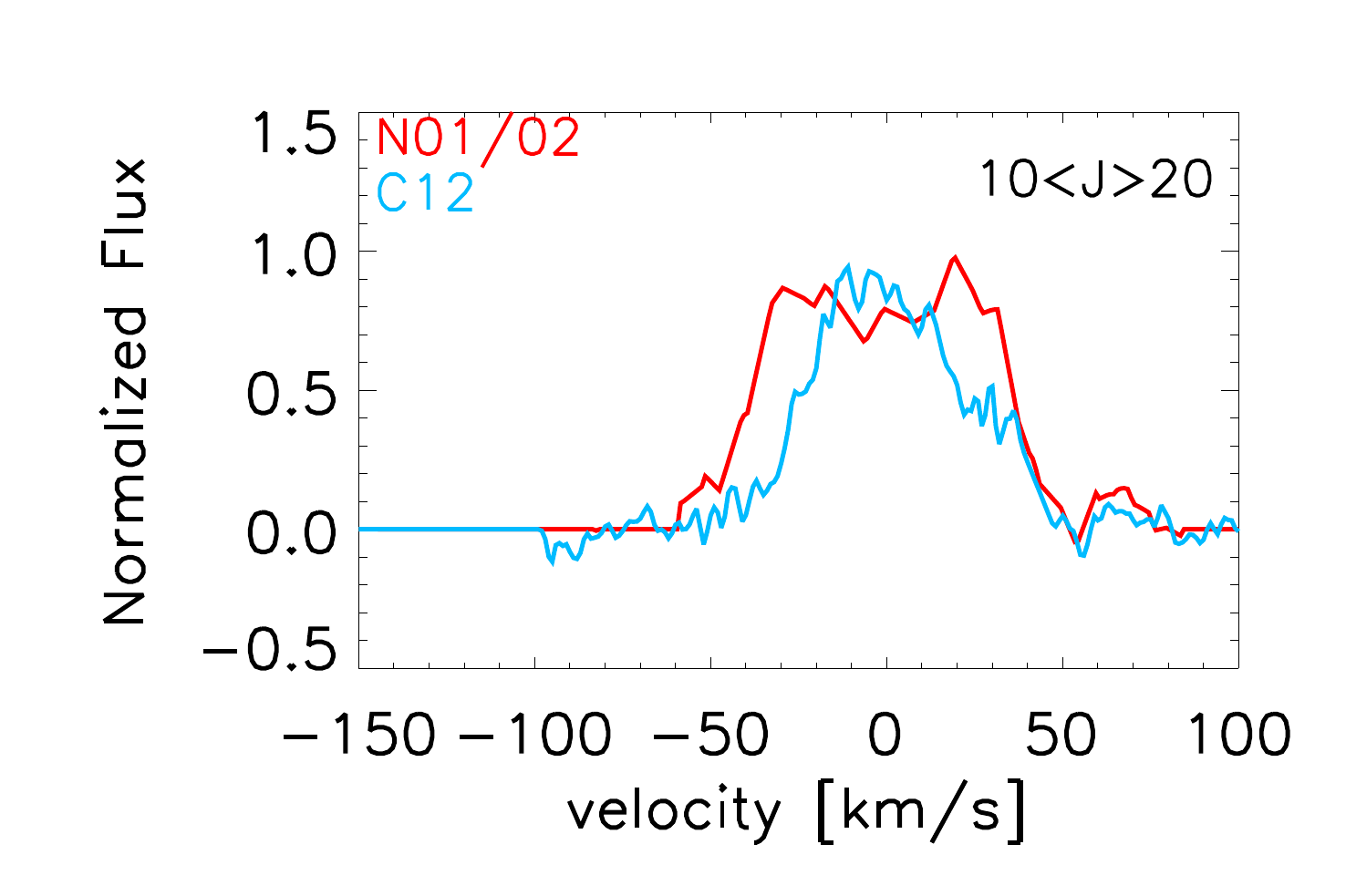}
\end{minipage}
\\
\begin{minipage}[r]{.36\textwidth}
  \includegraphics[width=\textwidth]{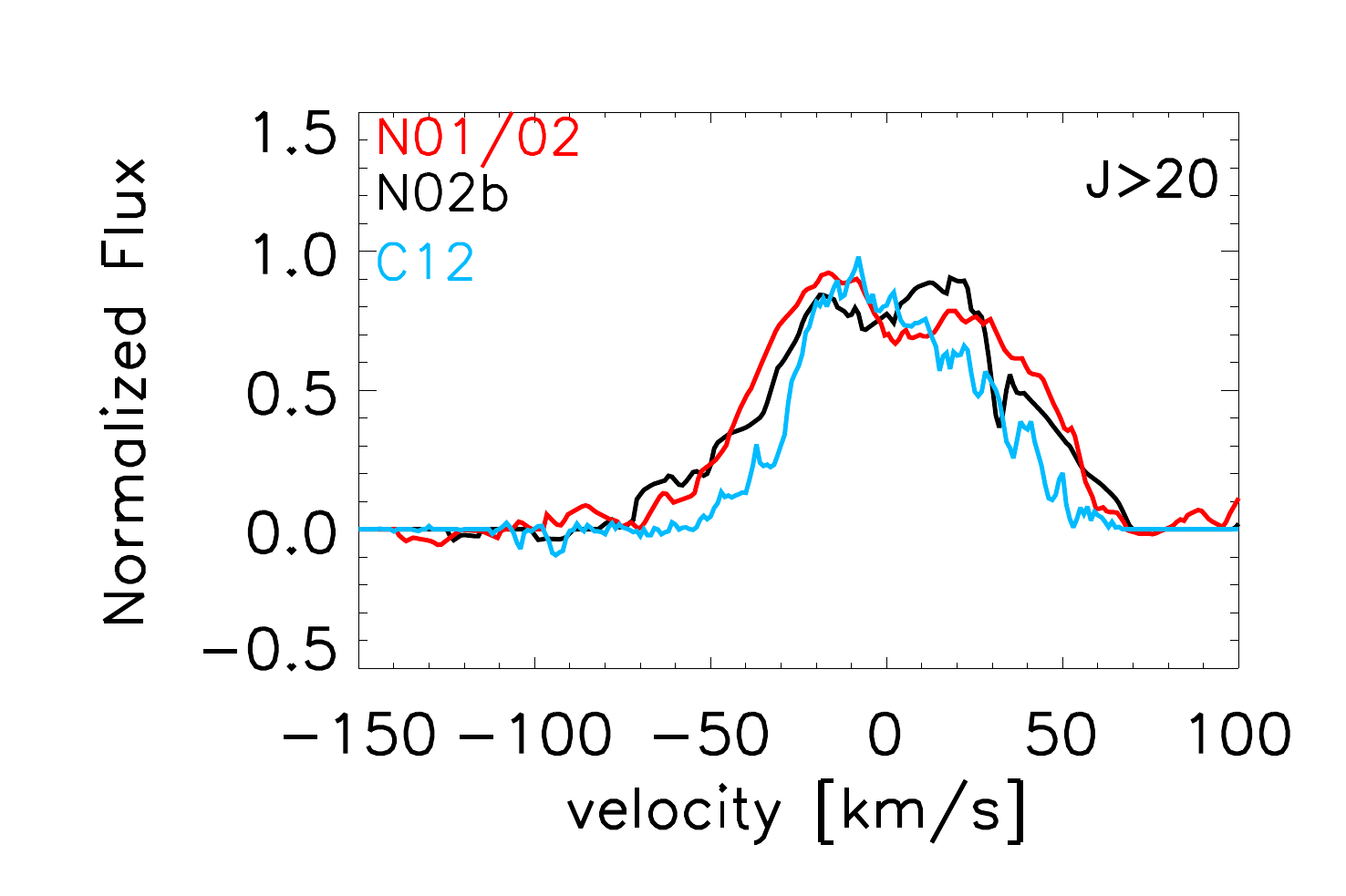}
\end{minipage}
\begin{minipage}[l]{.36\textwidth}
  \includegraphics[width=\textwidth]{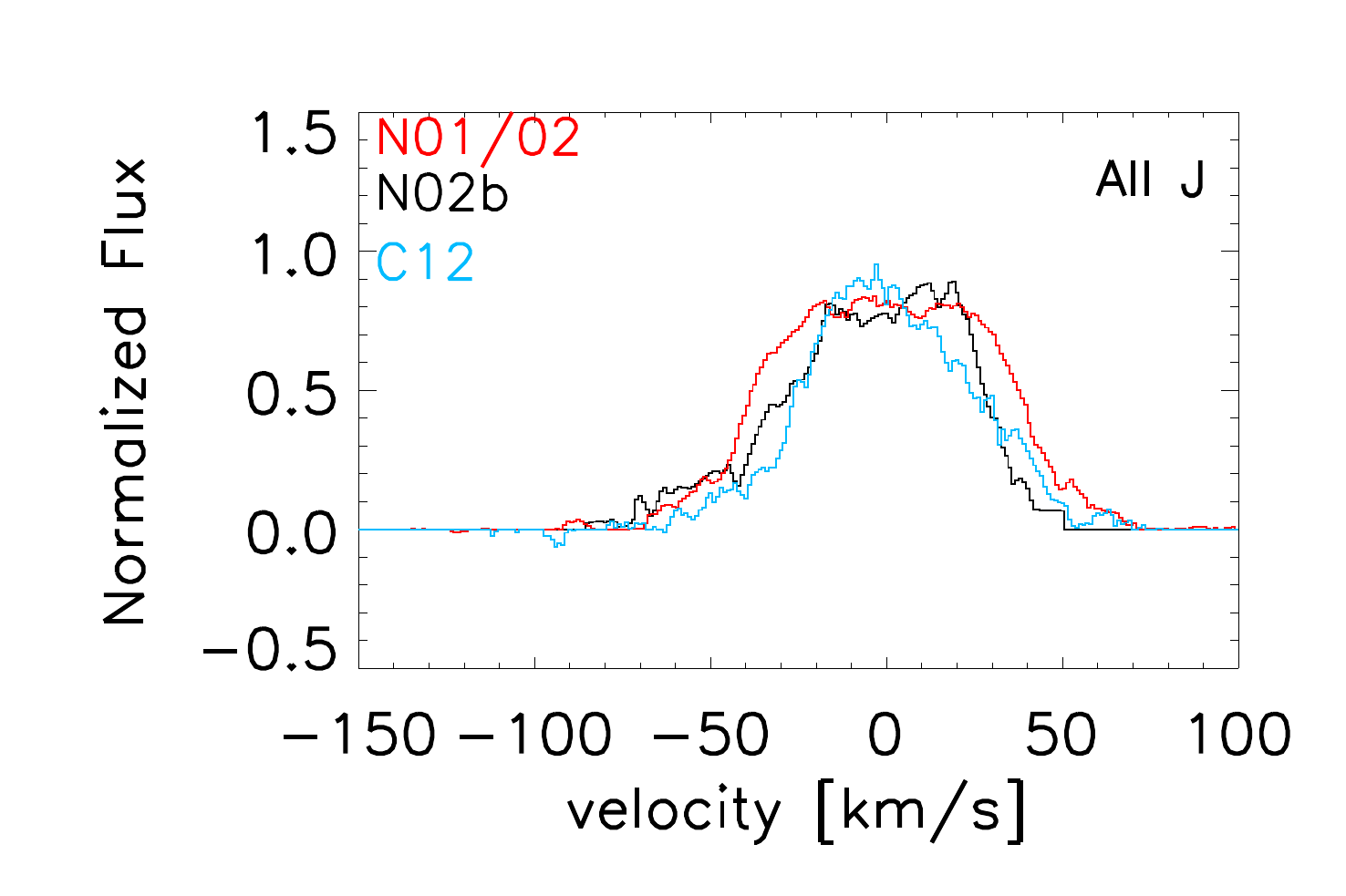}
\end{minipage}
\caption{Comparison of the median line profiles from the C12 data (blue), the N01/02 data (red), and the N02b data (black). Ranges of $J$-levels used for the medians (co-addition) are indicated on the plots.}
         \label{fig:median}
\end{figure*}

The individual lines collected from the three observational datasets are shown in Fig. \ref{fig:aalines}. For the CRIRES data the spectral resolution is $R=100000$, corresponding to $\Delta v$=3\,km/s, while for NIRSPEC it is $R=25000$, corresponding to $\Delta v$=12\,km/s. The lines displayed in Fig. \ref{fig:aalines} have smaller spacing between velocity channels than the $\Delta v$=3\,km/s and $\Delta v$=12\,km/s mentioned above (the C12 data has $\sim$1.5\,km/s, the N01/02 data has $\sim$4\,km/s, and the N02b data has $\sim$5\,km/s) due to fine sampling of the spectral resolution.  
In Fig. \ref{fig:median}, we show normalised line profiles compiled from co-adding \mbox{either} all lines, all low $J$ lines ($J$<10), all mid $J$ lines (10<$J$<20), or all high $J$ lines ($J$>20), in separate medians for C12, N01/02 and N02b.

\begin{table}
\caption{Observing details, FWHM and errors from Gaussian fits. N01/02= NIRSPEC 2001/2002 (Salyk et al. 2011), N02b= NIRSPEC 2002 (Brittain et al. 2007), C12= CRIRES 2012 (Hein Bertelsen et al. submitted).} 
%\tiny
\label{tab:fwhm}      % is used to refer this table in the text
\centering                          % used for centering table
\begin{tabular}{lllllll}        % centered columns (4 columns)
\hline            
\\
Dataset				& N01/02			&N02b		&C12		\\   
\hline            
Instrument		&NIRSPEC		&NIRSPEC&CRIRES\\
Observing dates	&06-08-2001		&23-03-2002	&06-03-2012	\\
			&08-08-2001		&	&			\\
			&21-04-2002	&	&			\\
			&22-07-2002	&	&			\\
$R$				&25.000			&25.000		&100.000		\\
$F_{\rm cont}$ at 4.770 $\mu$m&	12.1 Jy&---		&	16.3 Jy\\
Observed lines			&$J$=1--37	& $J$=1--33&$J$=0--36	\\
\hline                           
\\
FWHM				&[km/s]			&[km/s]		&[km/s]\\
\hline                           
all $J$		&73.9$\pm$1.4	&60.0$\pm$1.4&55.0$\pm$2.5\\   
$J$<10		&66.9$\pm$1.8	&45.8$\pm$1.6&54.5$\pm$3.9\\   
10<$J$<20	&74.3$\pm$3.8	&---			&50.9$\pm$4.9\\   
 $J$>20		&85.0$\pm$2.8	&78.9$\pm$2.2&59.2$\pm$4.3\\

\hline 
\end{tabular}
\end{table}

\begin{figure*}
\begin{center}$
\begin{array}{cc}
  \includegraphics[width=.45\textwidth]{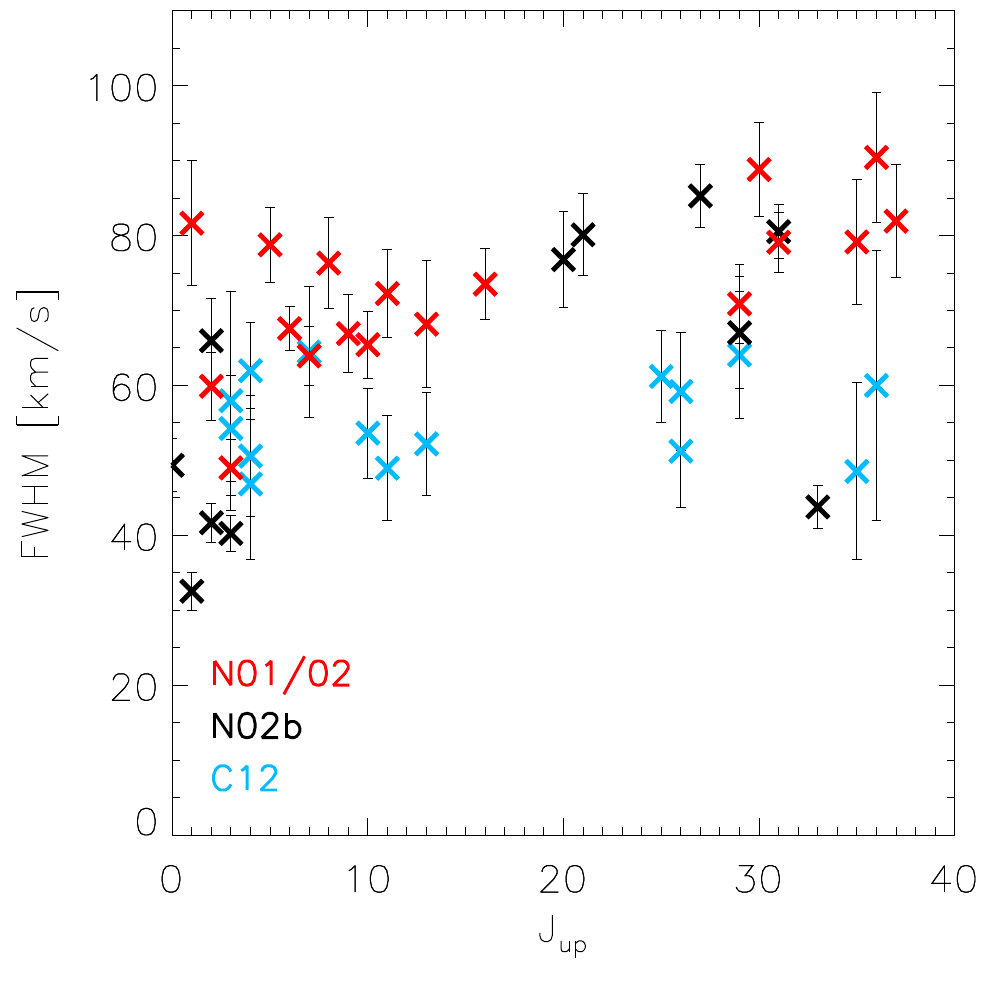}
\end{array}$
\end{center}
\caption{FWHM versus $J_{\rm up}$ for all three samples. Lines collected from C12 are shown in blue, lines collected from N01/02 are shown in red, and lines from N02b are shown in black.}
         \label{fig:fwhm}

\centering
\begin{minipage}[l]{.35\textwidth}
  \includegraphics[width=\textwidth]{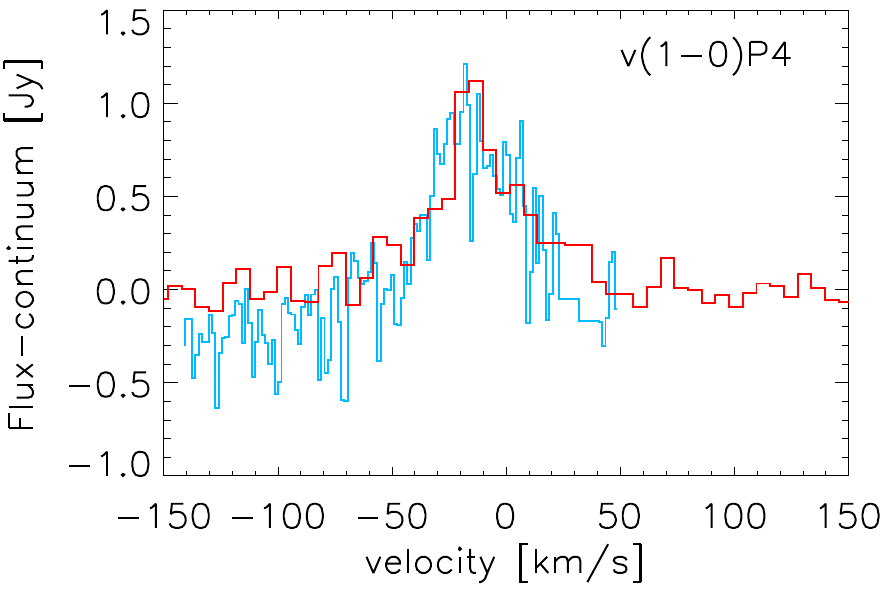}
\end{minipage}
\begin{minipage}[r]{.35\textwidth}
  \includegraphics[width=\textwidth]{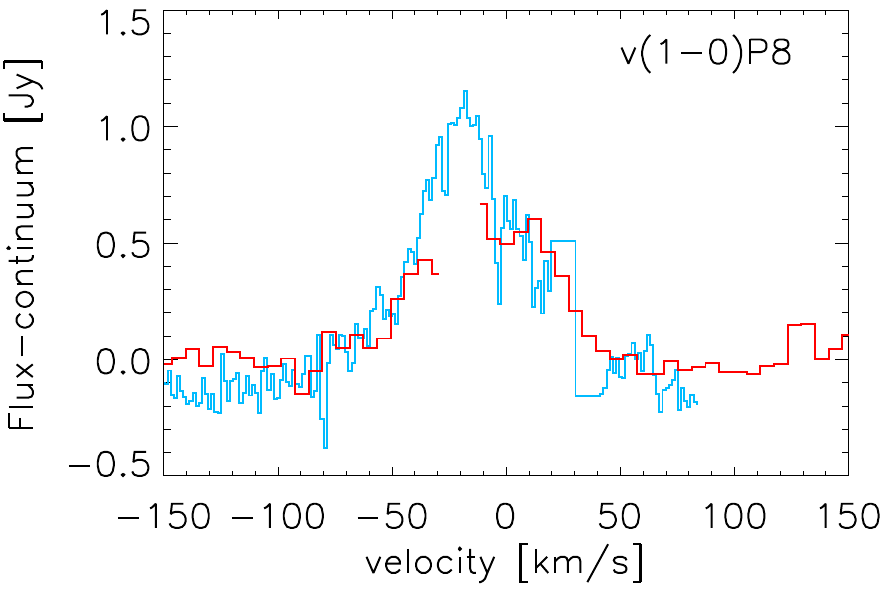}\\
\end{minipage}
\begin{minipage}[r]{.35\textwidth}
  \includegraphics[width=\textwidth]{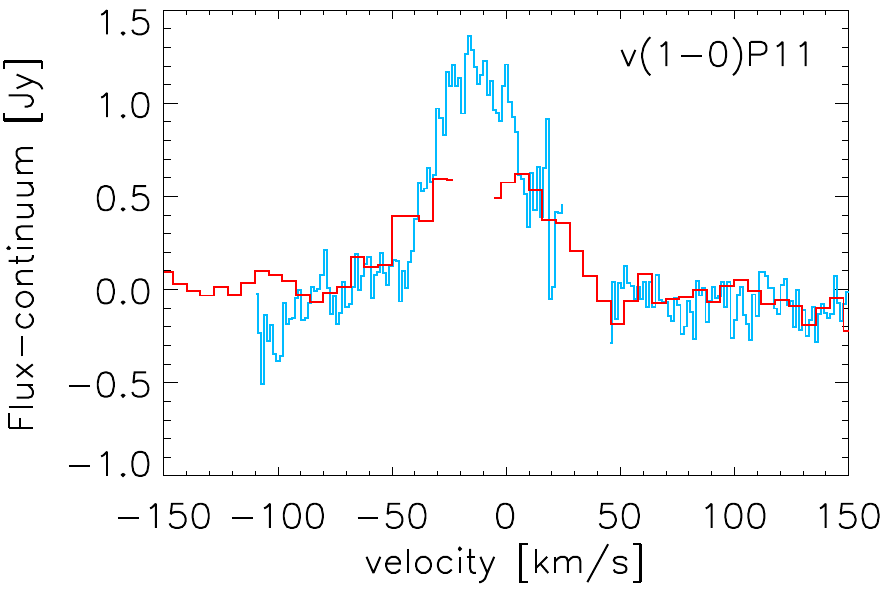}
\end{minipage}
\begin{minipage}[l]{.35\textwidth}
  \includegraphics[width=\textwidth]{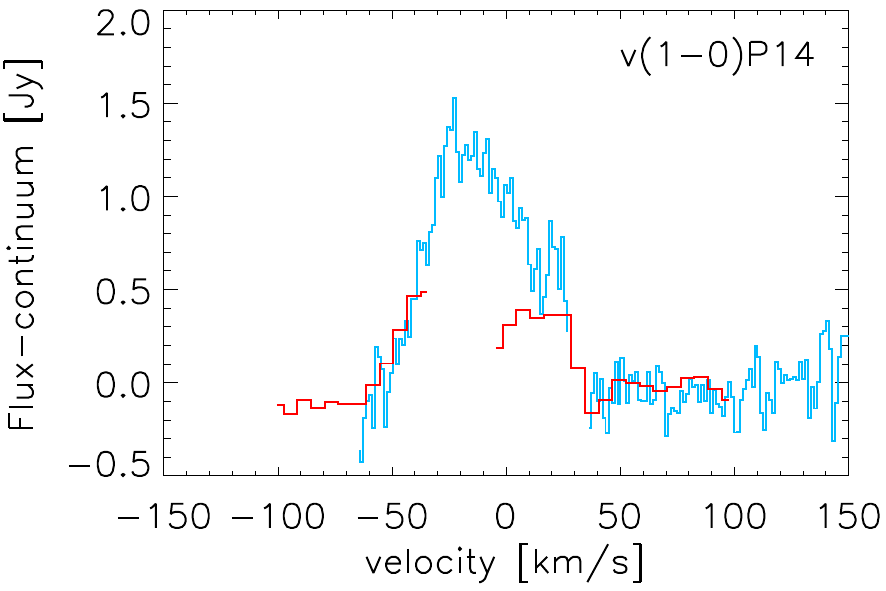}\\
\end{minipage}
\begin{minipage}[r]{.35\textwidth}
  \includegraphics[width=\textwidth]{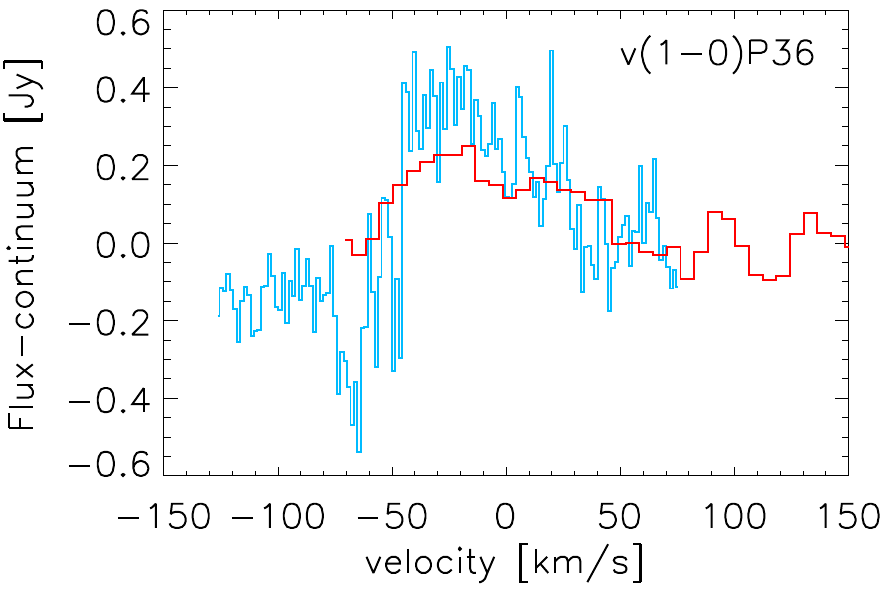}
\end{minipage}
\begin{minipage}[r]{.35\textwidth}
  \includegraphics[width=\textwidth]{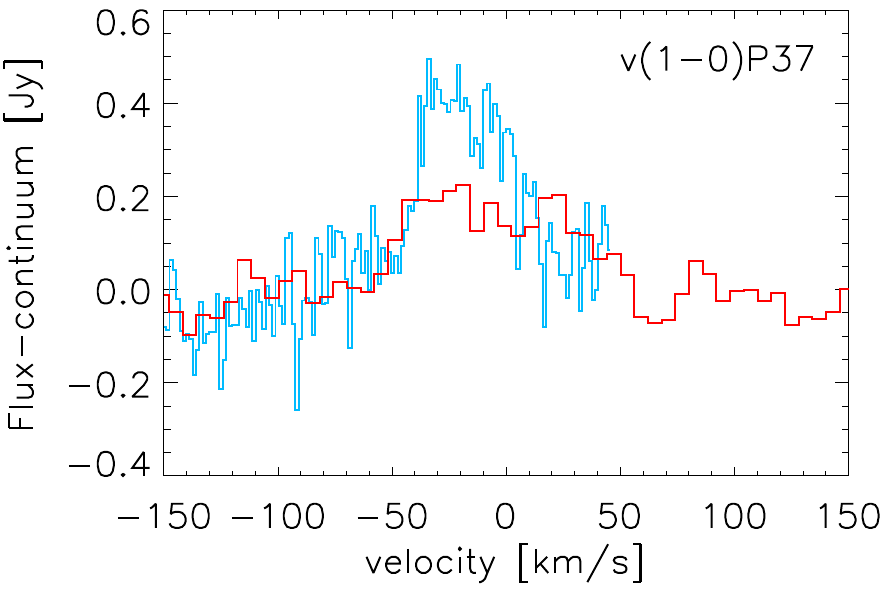}
\end{minipage}
\caption{Individual line profile comparisons of flux calibrated lines, that were present in both the C12 data (blue) and the N01/02 data (red). Line identifications are noted on the plots. A wide plotting range is chosen to show the typical noise on the continuum.}
         \label{fig:lineex}
\end{figure*}

The C12 spectrum shows single peaked emission lines for low and mid $J$ values while flat topped asymmetric profiles with shoulders seem to be present at higher $J$. The N01/02 spectrum shows single peaked emission lines for low $J$, while mid and high $J$ lines show either double peaked or flat topped profiles. However, the central dips in the double peaked profiles are comparable to the noise, and many double peaked profiles have central gaps in the spectra, where low transmission regions due to strong telluric absorption occurred (Fig. \ref{fig:aalines}). Even line profiles that display what looks like the normal central dip of a Keplerian profile could in fact be affected by telluric residuals that mimic the typical shape. Hence, the double peaked nature is uncertain for low and mid $J$ lines. High $J$ lines however, are unlikely to be affected heavily by telluric absorption (lines of $J \sim$25--30 or higher show no significant telluric absorption, se e.g. the C12 spectra), and are therefore more reliable. The N02b spectrum shows a general mix between single peaked and double peaked emission lines with no clear connection with high or low $J$. Also, several lines suffer from telluric overcorrection giving irregular lines. Again, the central dip is mostly comparable to the noise and a single peaked nature cannot be ruled out from this dataset either. CRIRES has higher spectral resolution than NIRSPEC, and the C12 data have many high signal to noise lines. Hence, if present at the time of observation, the double peak should have been visible in this spectrum. 

However, the C12 dataset has been collected a decade later than the other two datasets, thus it is possible that the variability noted for this source in the near-IR continuum and in the optical \citep{sitko2008, ellerbroek2014} plays a role.
For these studies observations collected at several different epochs were used dating from 1979 until 2012. During this time, the NIR brightness was increased by more than 10\% (in $H$-, $K$-, $L$-, and $M$-band) in 3 epochs: 1986, 2001--2002, and 2011--2012. 
Each of the CO ro-vibrational datasets are collected during or shortly after one of these epochs (N01/02 in 2001--2002, N02b in 2002, and C12 in 2012).
Hence the observed variations of the CO ro-vibrational line profiles are not directly linked to the increased NIR brightness epochs. Meanwhile, this does not exclude that one underlying mechanism could in fact be driving both the NIR variability and the CO ro-vibrational variability in different ways.

From all three datasets, we find broad emission lines (FWHM$\sim$50--80\,km/s) suggesting that the CO ro-vibrational lines are emitted from the inner radius of the disc. 
However, the FWHM measured from the three sets are not in agreement. In Fig. \ref{fig:fwhm}, we show the FWHM versus $J_{\rm up}$ for lines collected from all three datasets in the same manner. The FWHM variations can also be seen from the medians collected for high, mid and low $J$ separately, displayed in Fig. \ref{fig:median} and listed with FWHM in Table \ref{tab:fwhm}. We see that lines from N01/02 are overall wider than those from C12. Low $J$ lines from N02b are narrower than both N01/02b and C12, but high $J$ lines from N02b are comparable in FWHM to N01/02. As discussed above, the N02b dataset has several irregular lines, which may be due to imperfect telluric corrections. Thus for the further comparison we focus on the C12 and N01/02 datasets. 

To explore in more details the differences between C12 and N01/02, we also compare flux calibrated line profiles for individual transitions that were present in both data sets (Fig. \ref{fig:lineex}). This shows that the general width and line wings of the individual line profiles are the same in the two data sets, while (in the case of  high and medium $J$ lines) the intensity and the shape of the central part of the lines differ. For low $J$ (the P4 line) the shape is well matched in the two datasets. An interpretation could be that the single peaked line profiles (i.e. lines from the CRIRES data and low $J$ lines from the NIRSPEC data) are composites of multiple emitting regions, where the main (Keplerian) components are the same in the two epochs and the additional component, present in lines from C12 and in low $J$ lines from N01/02, could be related to the NIR variability. This additional component would then 'drive' the FWHM in C12 to lower values, since the half maximum location is shifted upwards.

\section{Comparison with a disc model} \label{sec:mod}
We use a previously published disc structure for HD\,163296 produced with ProDiMo {\citep[Model 3 from][]{Tilling2012}}, to model for the first time a sample of CO ro-vibrational lines. 
In the following sections (Sections \ref{sec:mod_desc} and \ref{sec:modres})we first describe the model and then present how the model results compare to the observed lines.

\subsection{Model description} \label{sec:mod_desc}
ProDiMo is a radiation thermo-chemical disc code \citep{woitke2009, kamp2010}, which solves the radiative transfer, the chemical network and the {gas heating/cooling balance.} The \citet{Tilling2012} model assumes a parameterised disc structure using power laws for the surface density and the gas scale height. Level populations are calculated from statistical equilibrium and detailed line transfer calculations are performed to predict emission lines. In order to model CO ro-vibrational emission lines, we use the large CO ro-vibrational model described in \citet{thi2012}. We include fluorescence pumping to the $A^{1}\Pi$ electronic level and use 40 rotational levels within 7 vibrational levels of both the ground electronic state $X^{1}\Sigma^{+}$ and the excited state $A^{1}\Pi$. 

The \citet{Tilling2012} model has been computed using a Monte-Carlo evolutionary $\chi^2$-minimisation strategy \citep{woitke2011}, varying 11 parameters to find the best fit to observed emission lines (not including the CO ro-vibrational lines) and the dust spectral energy distribution (SED), from HD\,163296, simultaneously. The spatial CO interferometric data were not used as a constraint for the modelling, however, the extracted line profiles for the pure rotational CO lines \citep{Isella2007} were used as a constraint for the radial extent and tapered edge.
\begin{figure*}
\begin{center}$
\begin{array}{cc}
  \includegraphics[width=.327\textwidth]{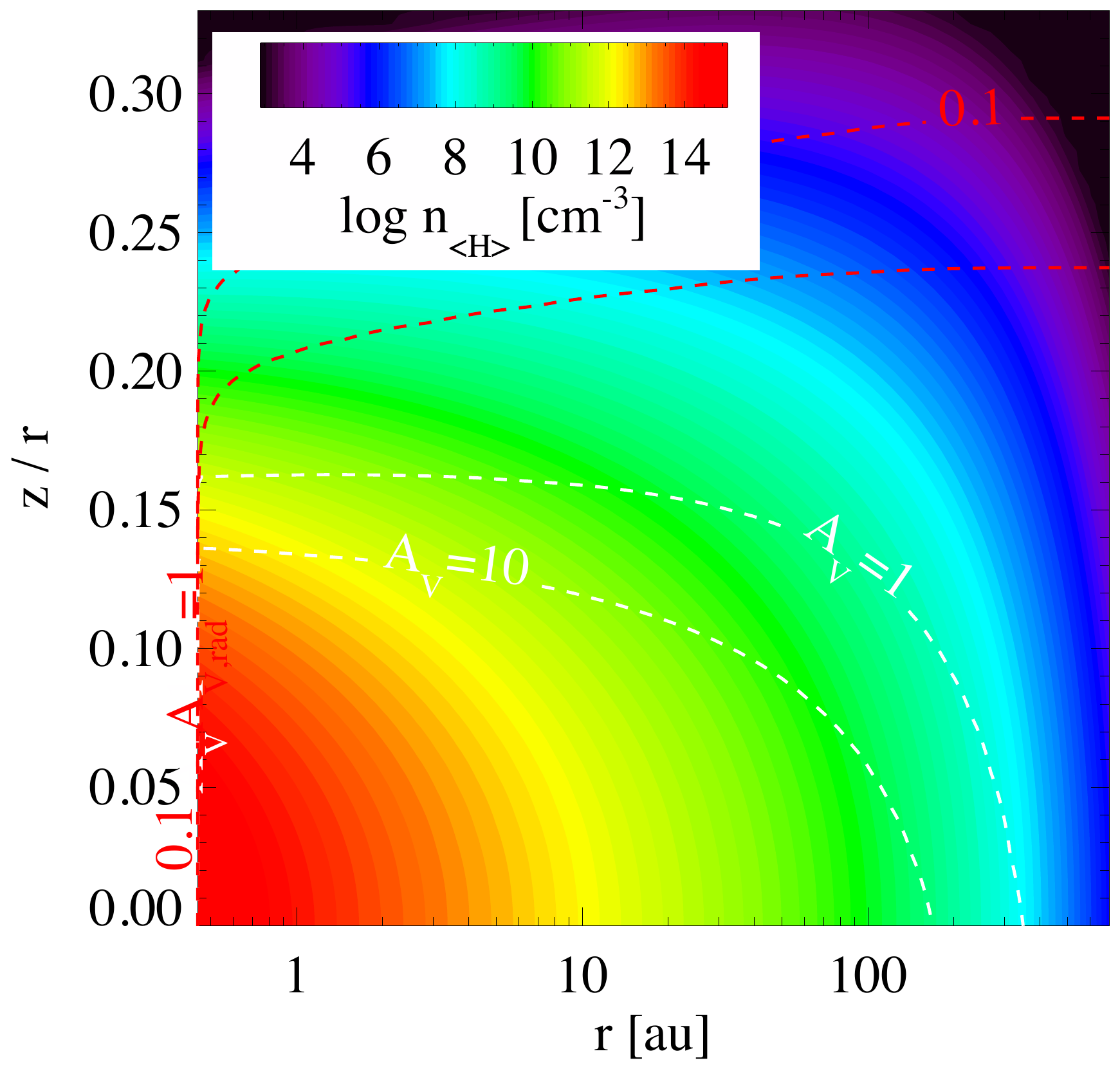}
  \includegraphics[width=.327\textwidth]{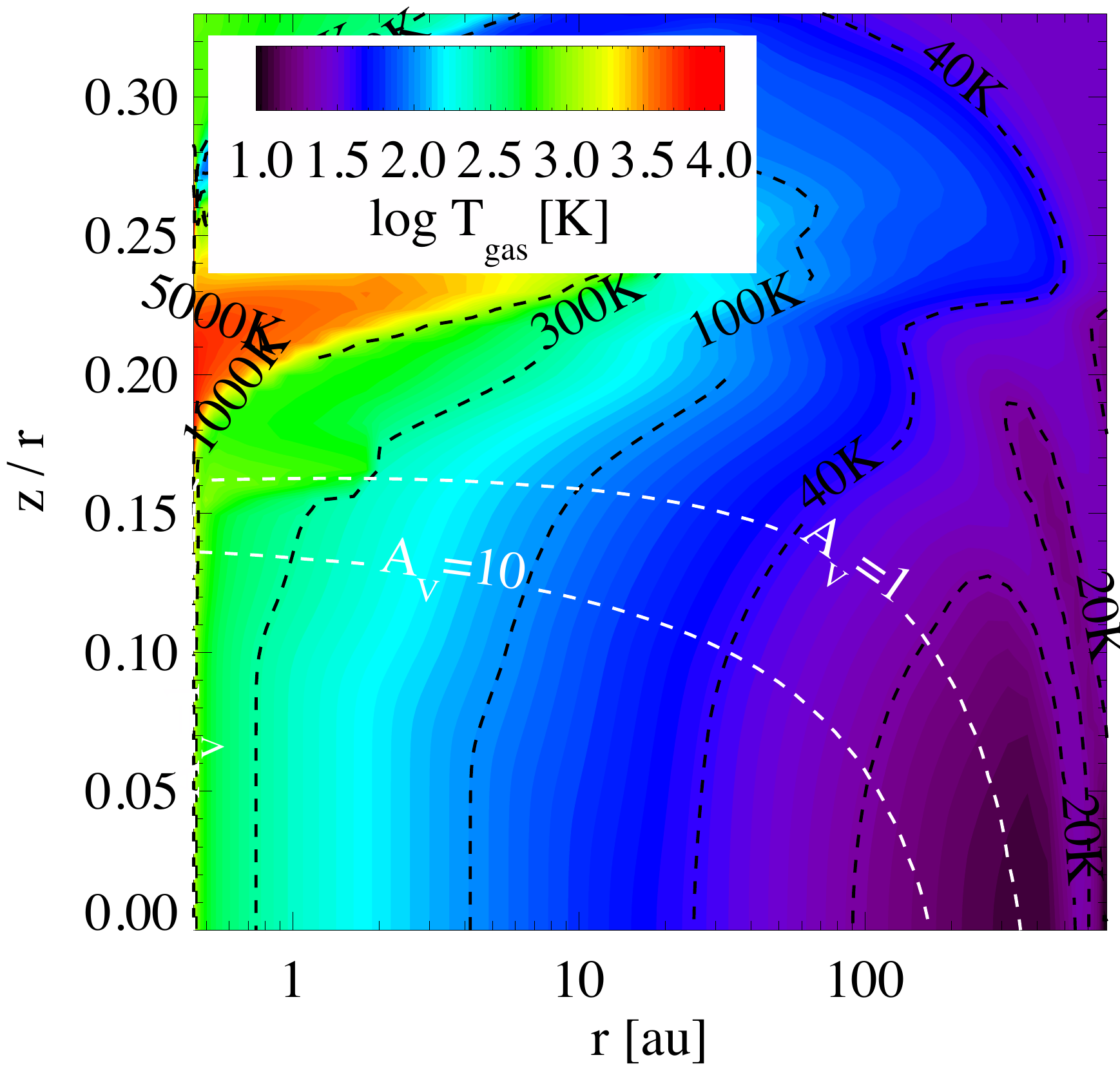}
  \includegraphics[width=.327\textwidth]{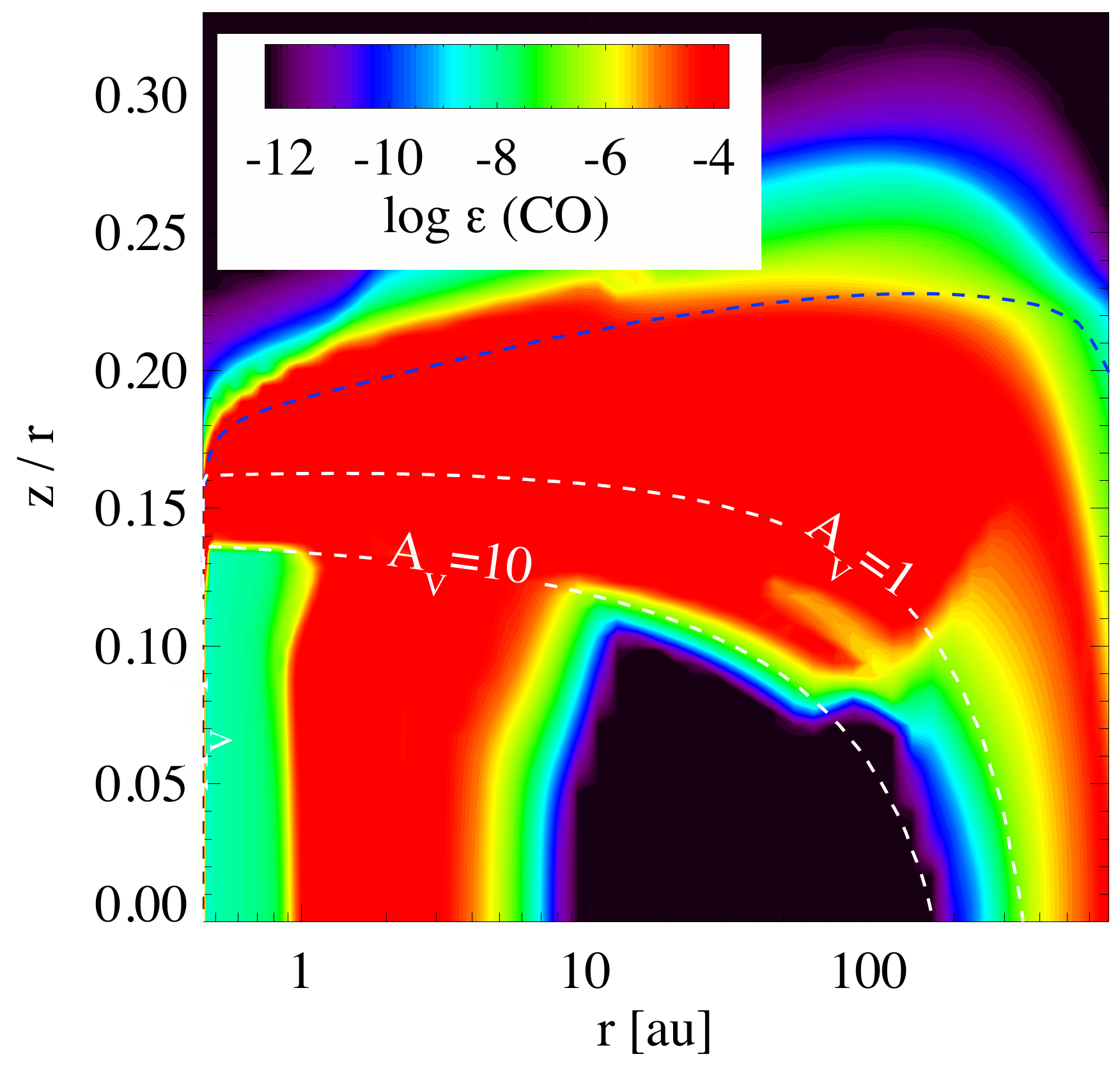}\\
 \end{array}$
\end{center}
\caption{Left panel: Gas volume density, together with contours showing $A_{\rm v}$=1.0 and $A_{\rm v}$=10.0 (dashed white lines), $\rm{A_{v,rad}}$=0.1, and $\rm{A_{v,rad}}$=1.0 (dashed red lines). Middle panel: Gas temperature, together with contours showing $A_{\rm v}$=1.0 and $A_{\rm v}$=10.0 (white dashed lines), and gas temperature contours at 20, 40, 100, 300 and 1000\,K (black dashed lines). Right panel: CO abundance, together with contours showing $A_{\rm v}$=1.0 and $A_{\rm v}$=10.0 (dashed white lines) and log$(\chi/n)$=-3 (dashed blue line). }
         \label{fig:mod_nh_a}
\end{figure*}
\begin{table}
\caption{\{{Key model parameters adapted or fitted in T12 or adapted from other work: B10=\citet{benisty2010}, B16=this work, E09=\citet{eisner2009}, G00=\citet{grady2000}, H08=\citet{hughes2008}, I07=\citet{Isella2007}, L07=\citet{leeuwen2007}, M09=\citet{montesinos2009}, R10=\citet{renard2010}, T08a=\citet{tannirkulam2008}, T08b=\citet{tannirkulam2008b}, T12=\citet{Tilling2012}, W06=\citet{wassell2006}. For more details see T12.}}  % title of Table
%\tiny
\label{tab:mod_hd163296}      % is used to refer this table in the text
\centering                          % used for centering table
\begin{tabular}{lrrrrrr}        % centered columns (4 columns)
\hline            

%						&\#1 \\
Quantity						&Symbol   &Value&  Reference\\
\hline                           
 inner radius&$R_{\rm in}$[au]				&{0.45}			&E09,T08b,\\
			&							&				&R10,B10\\
 outer radius& $R_{\rm out}$ [au]				&700.0			&B16\\
 tapering-off radius &$R_{\rm taper}$ [au]		&{125.}			&H08\\
stellar luminosity &$L_{*}/L_{\astrosun}$		&37.7			&M09\\
stellar mass&$M_{*}/M_{\astrosun}$ 			&2.47			&M09\\
effective temperature&$T_{\rm eff}$[K]			&9250.0			&M09\\
disc gas mass&$M_{\rm disc}/M_{\astrosun}$		&0.071			&Fit in T12\\
dust to gas mass ratio&dust-to-gas			&0.01			&Fit in T12	\\
dust grain material -\\
 mass density&$\rho_{\rm grain}$[g/cm$^3$]	&3.36			&T12\\
minimum grain size&$a_{\rm min}$[$\mu$m]	&0.01			&Fit in T12\\
maximum grain size&$a_{\rm max}$[$\mu$m]	&2041			&Fit in T12\\
dust size distribution -\\
power law index&$p$						&3.68			&Fit in T12\\
flaring index&$\beta$						&1.066			&Fit in T12\\
distance &$d$ [pc]							&118.6			&L07\\
PAH abundance -\\
relative to ISM&$f_{\rm{PAH}}$				&{0.0068}			&Fit in T12\\
inclination&$i$								&50.0$\degree$	&G00,W06,I07,\\
			&							&				& T08a,B10\\
\hline 
\end{tabular}

\hspace{50 mm}

\caption{CO line fluxes. The model line fluxes are computed from escape probability and high $J$ rotational lines were observed with Herschel/PACS \citep{Tilling2012}, low $J$ rotational lines were observed from the ground \citep{Isella2007}, and the observed ro-vibrational line fluxes are from \citet{salyk2011}.}             % title of Table
%\tiny
\label{tab:mod_allobs}      % is used to refer this table in the text
\centering                          % used for centering table
\begin{tabular}{lllllll}        % centered columns (4 columns)
\hline            
	Mol.		&$\lambda$	&Obs.	&Model \\
			&	[$\mu$m]&[10$^{-18}$W/m$^2$]	&[10$^{-18}$W/m$^2$]	\\
\hline                          
$^{13}$CO $J$=1--0&2720.41 &0.0124		&0.0105\\
CO $J$=2--1&1300.4 		&0.379		&0.40\\
CO $J$=3--2&866.96 		&1.65		&1.21\\ 
CO $J$=18--17&144.78		&<13.1		&2.69\\   
CO $J$=29--28&90.16		&<11.1		&0.624\\   
CO $v$=1--0 P37&5.053			&60.3		&122.\\
CO $v$=1--0 P14&4.793			&86.5.		&71.3\\
CO $v$=1--0 P4&4.700			&158		&70.8\\
\hline                           
\end{tabular}
\end{table}
Key parameters used in the model are displayed in Table \ref{tab:mod_hd163296}. The gas volume density distribution, the gas temperature and the CO abundance are shown in Fig. \ref{fig:mod_nh_a}.
In Table \ref{tab:mod_allobs}, emission line predictions from the model are compared to the observed lines from the literature. Even though this model has not been made to fit CO ro-vibrational observations, we use it here to investigate the emitting region of CO ro-vibrational transitions and to compare with observations.

\begin{figure*}
\centering
\begin{minipage}[r]{.32\textwidth}
   \includegraphics[width=\textwidth]{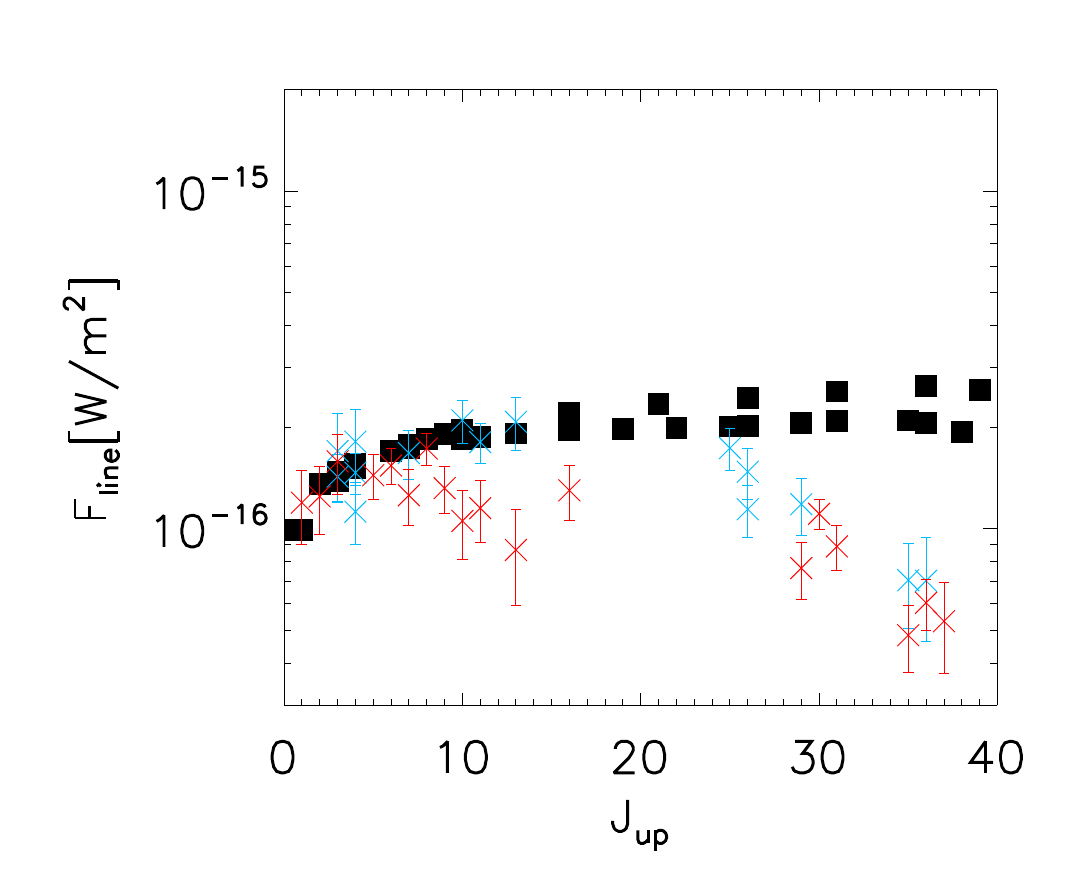}
\end{minipage}
\begin{minipage}[l]{.32\textwidth}
  \includegraphics[width=\textwidth]{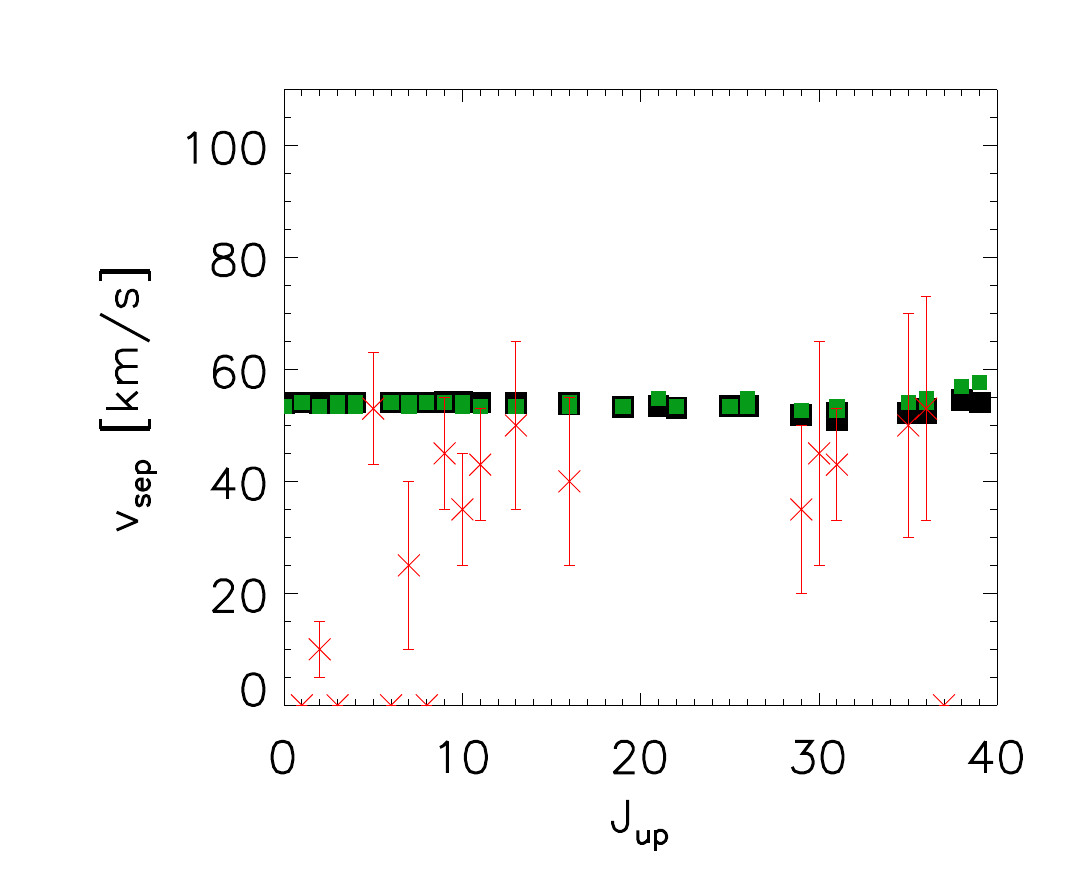}
\end{minipage}
\begin{minipage}[r]{.32\textwidth}
   \includegraphics[width=\textwidth]{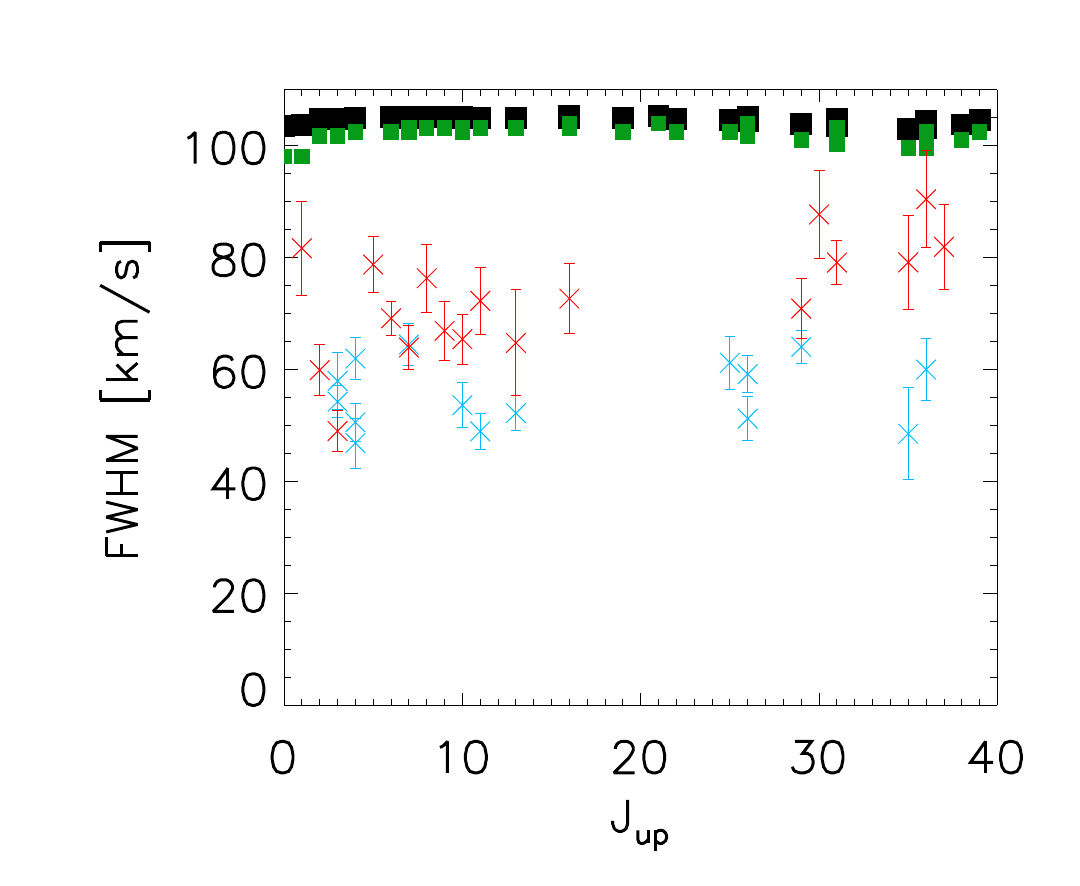}\\
\end{minipage}
\begin{minipage}[r]{.32\textwidth}
   \includegraphics[width=\textwidth]{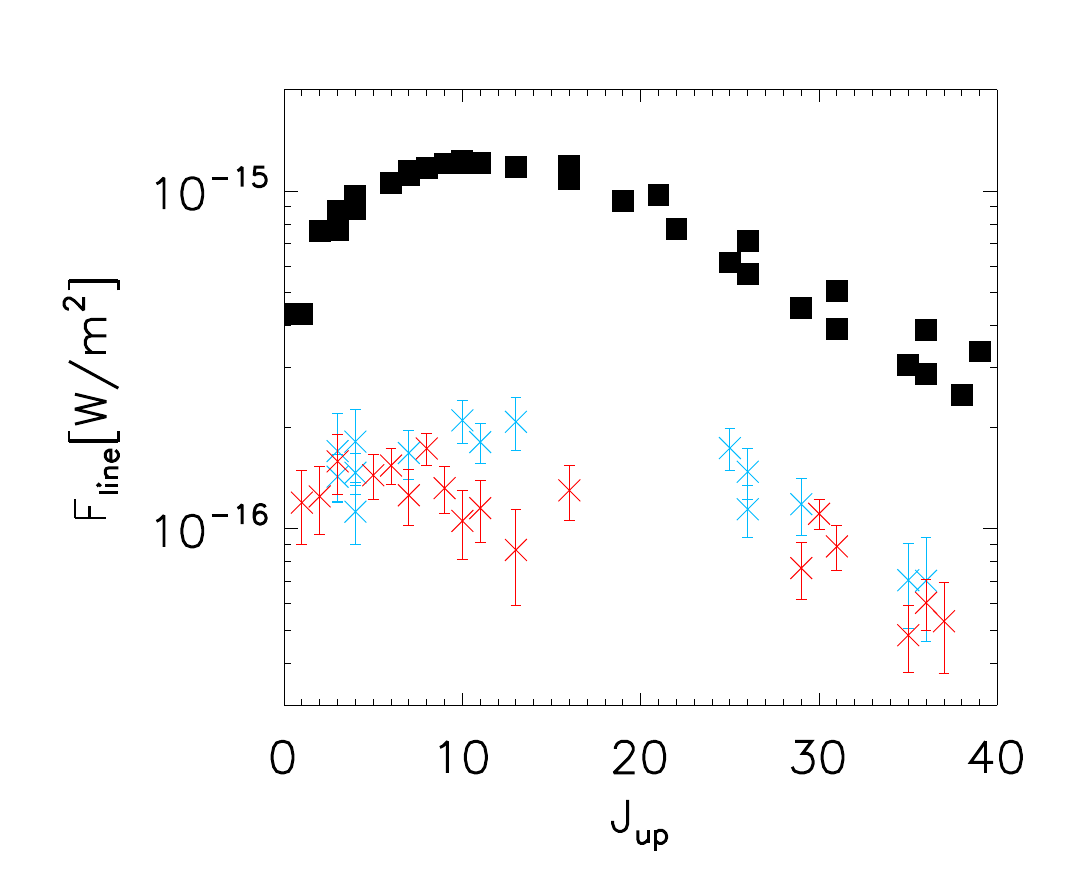}
\end{minipage}
\begin{minipage}[l]{.32\textwidth}
  \includegraphics[width=\textwidth]{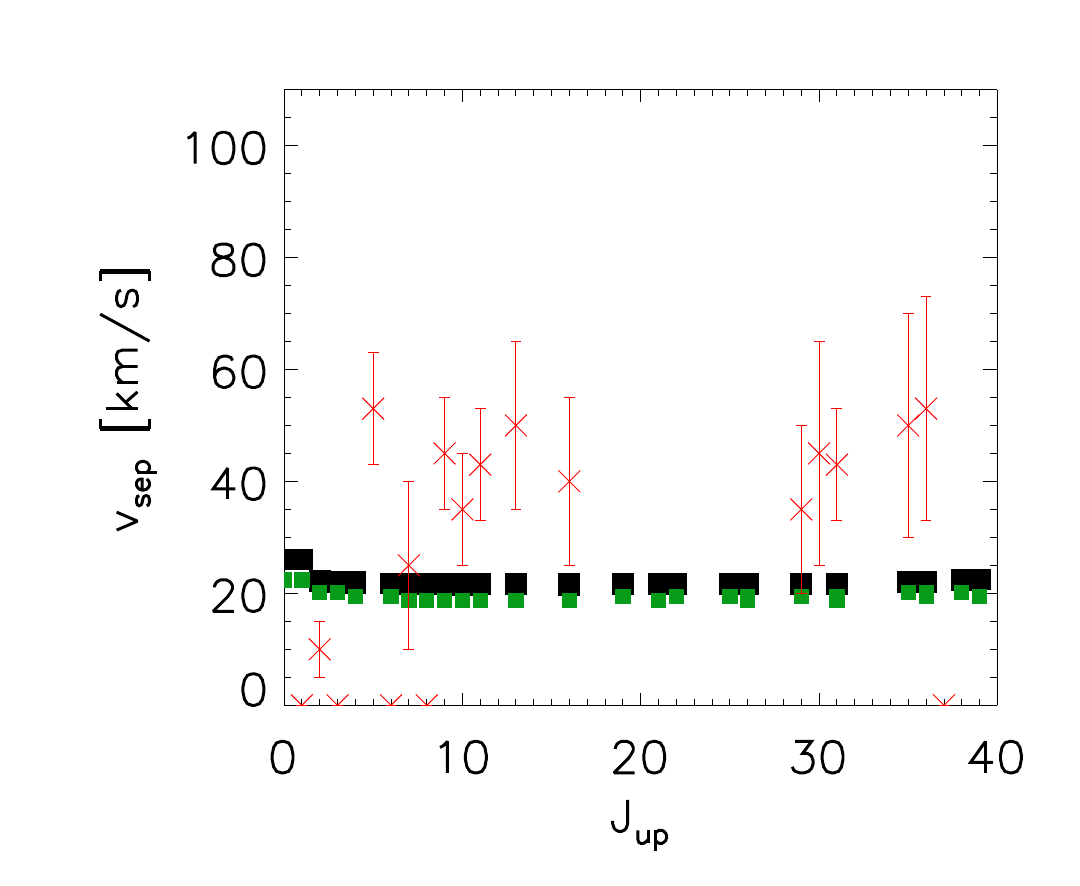}
\end{minipage}
\begin{minipage}[r]{.32\textwidth}
   \includegraphics[width=\textwidth]{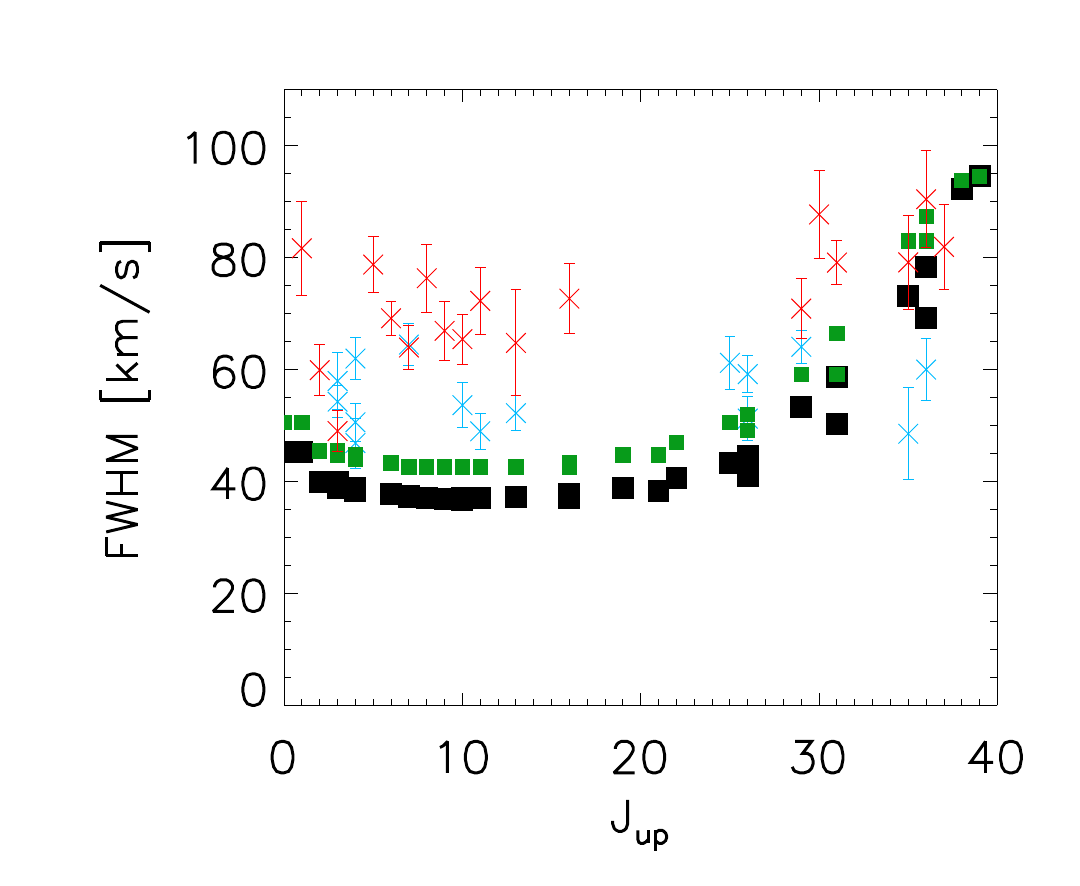}
\end{minipage}
\caption{Upper panels: The line fluxes (\textit{left}), the peak separations (\textit{middle}), and the FWHM (\textit{right}), as a function of $J_{up}$ for the modelled sample of CO fundamental ro-vibrational (black squares). The green squares are the model convolved with the NIRSPEC resolution (the model convolved with the CRIRES resolution corresponds very closely to the un-convolved model and is therefore not shown). Observations are shown for comparison as blue crosses (C12) and red crosses (N01/02). In addition to the displayed error bars for the observed lines, there is also an additional systematic error of 50\% for the C12 data and 20\% for the N01/02 data. This error has a similar effect on all lines, and thus do not affect line ratios. Bottom panels: The same plots with the lines calculated in LTE.} 
         \label{fig:mod}
\end{figure*}

\begin{figure*}
%\begin{figure*}
\centering
\begin{minipage}[l]{.32\textwidth}
\includegraphics[width=\textwidth]{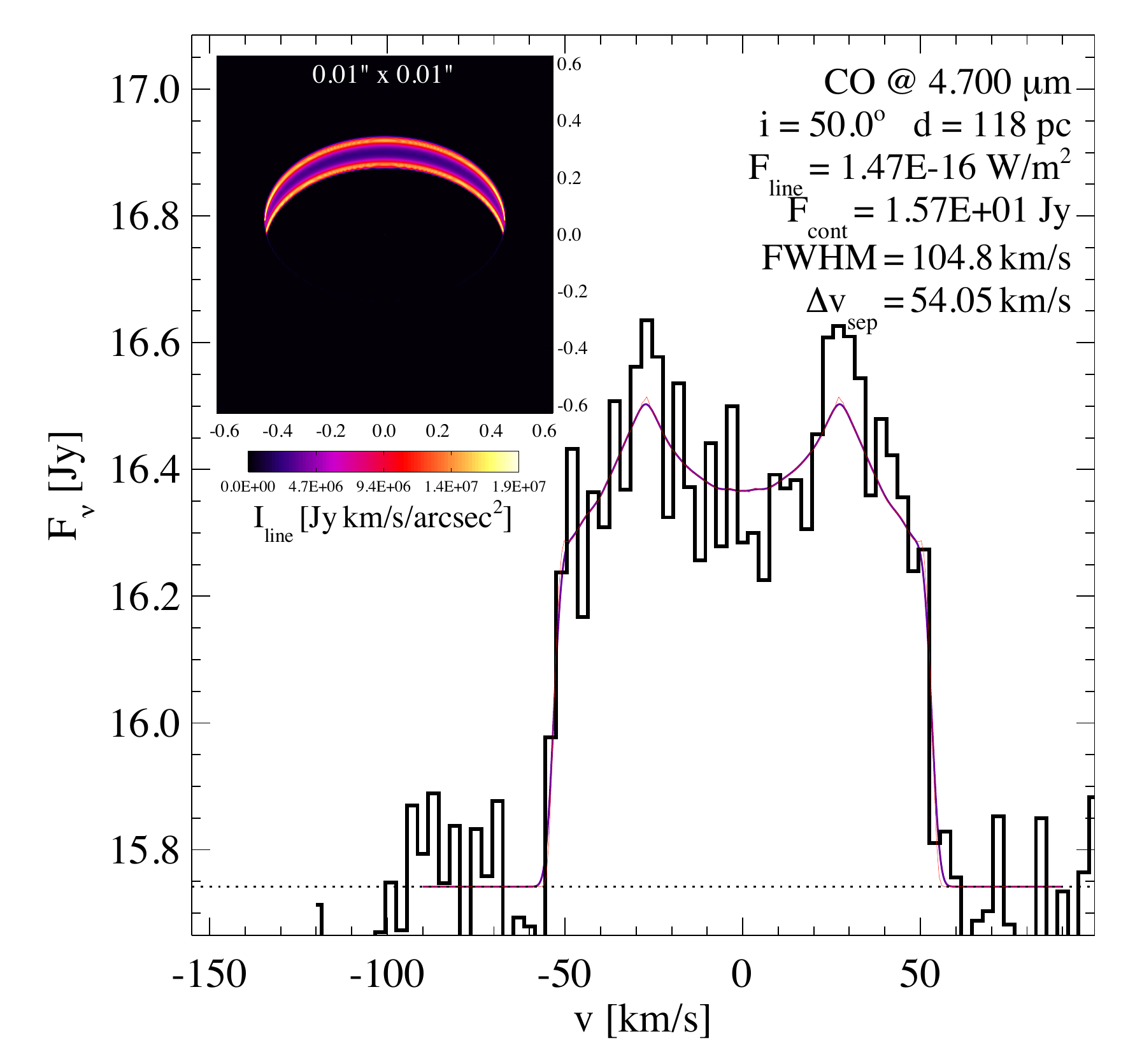}
\end{minipage}
\begin{minipage}[r]{.32\textwidth}
\includegraphics[width=\textwidth]{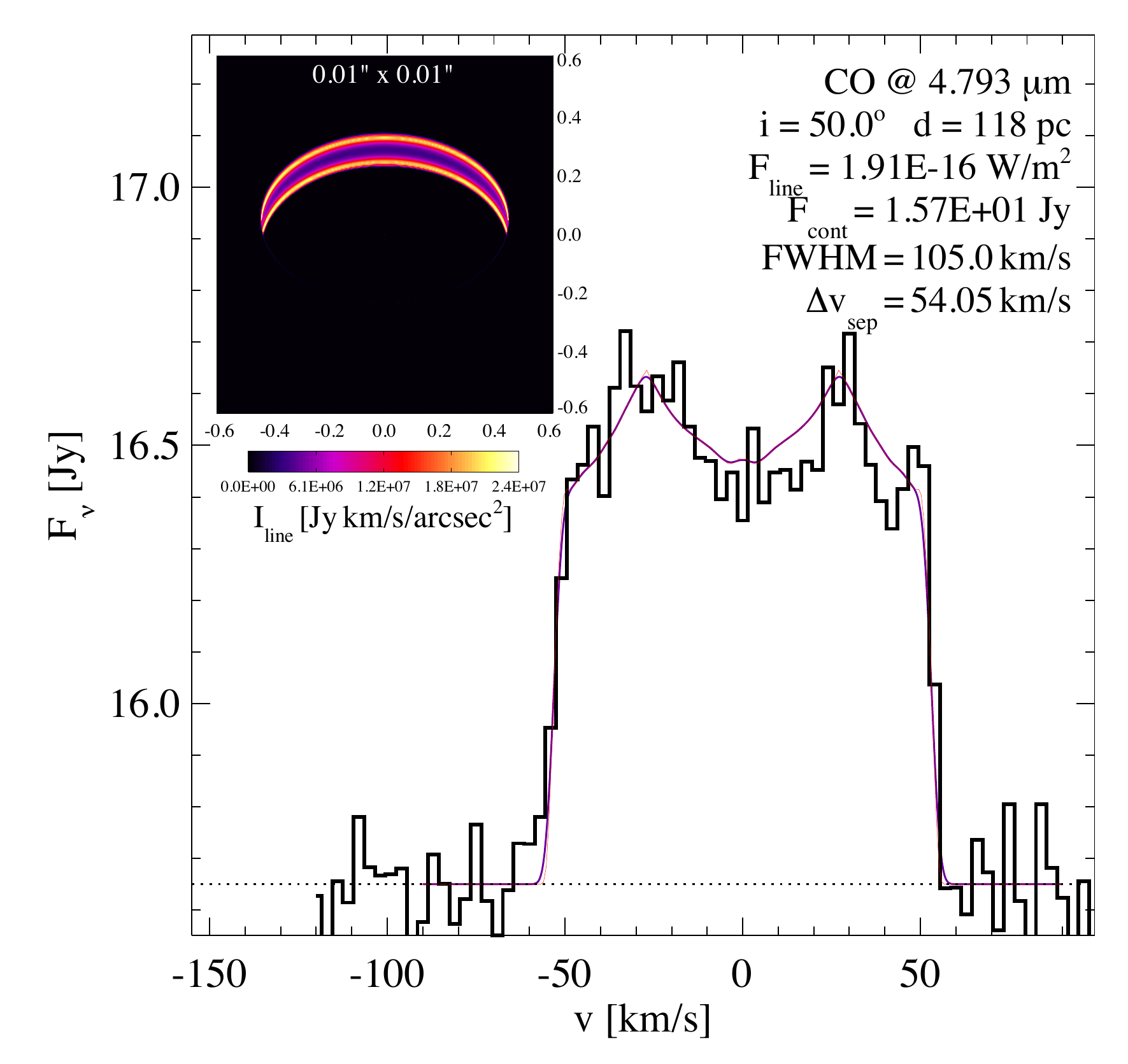}
\end{minipage}
\begin{minipage}[r]{.32\textwidth}
\includegraphics[width=\textwidth]{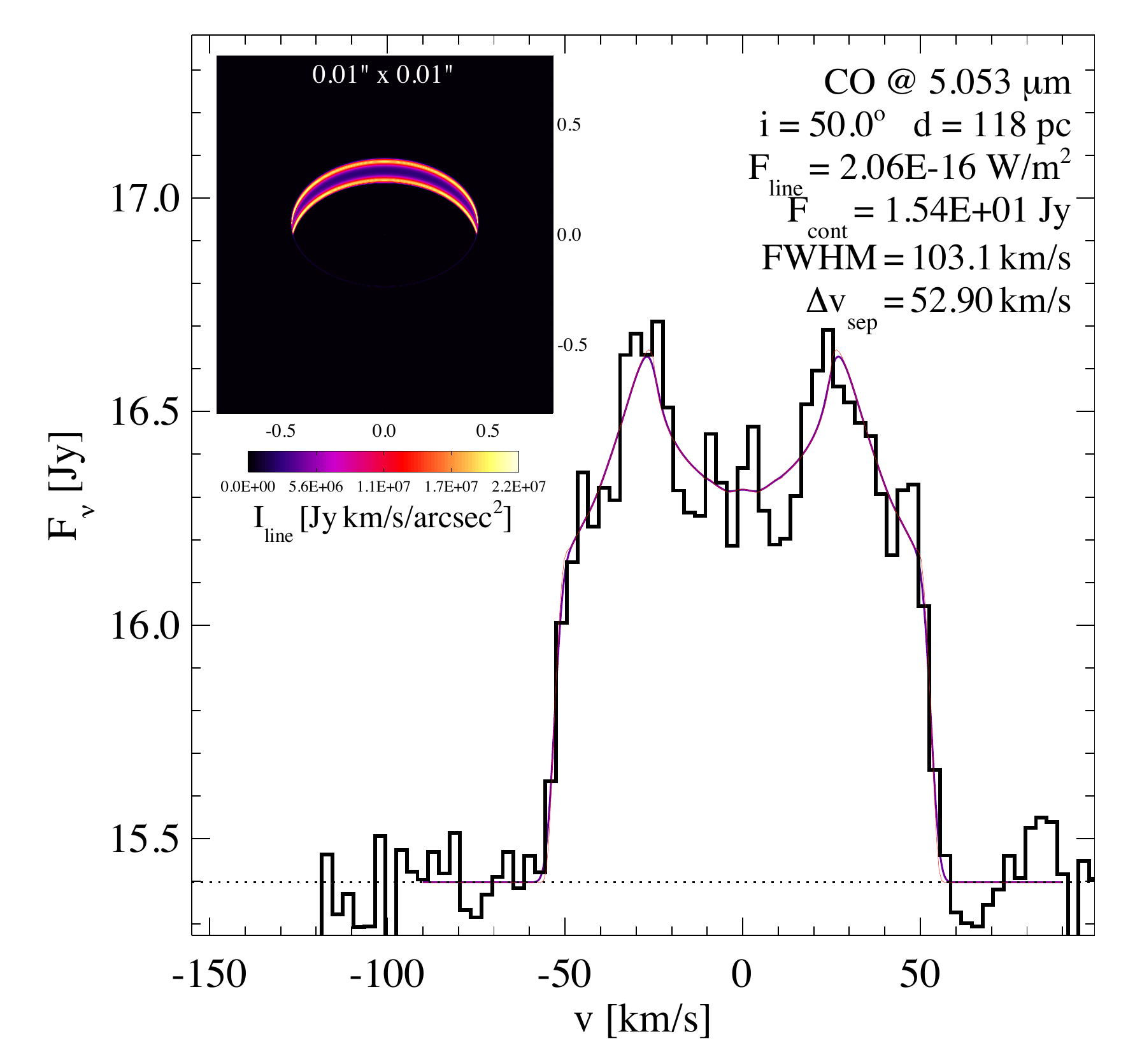}
\end{minipage}
\begin{minipage}[l]{.32\textwidth}
\includegraphics[width=\textwidth]{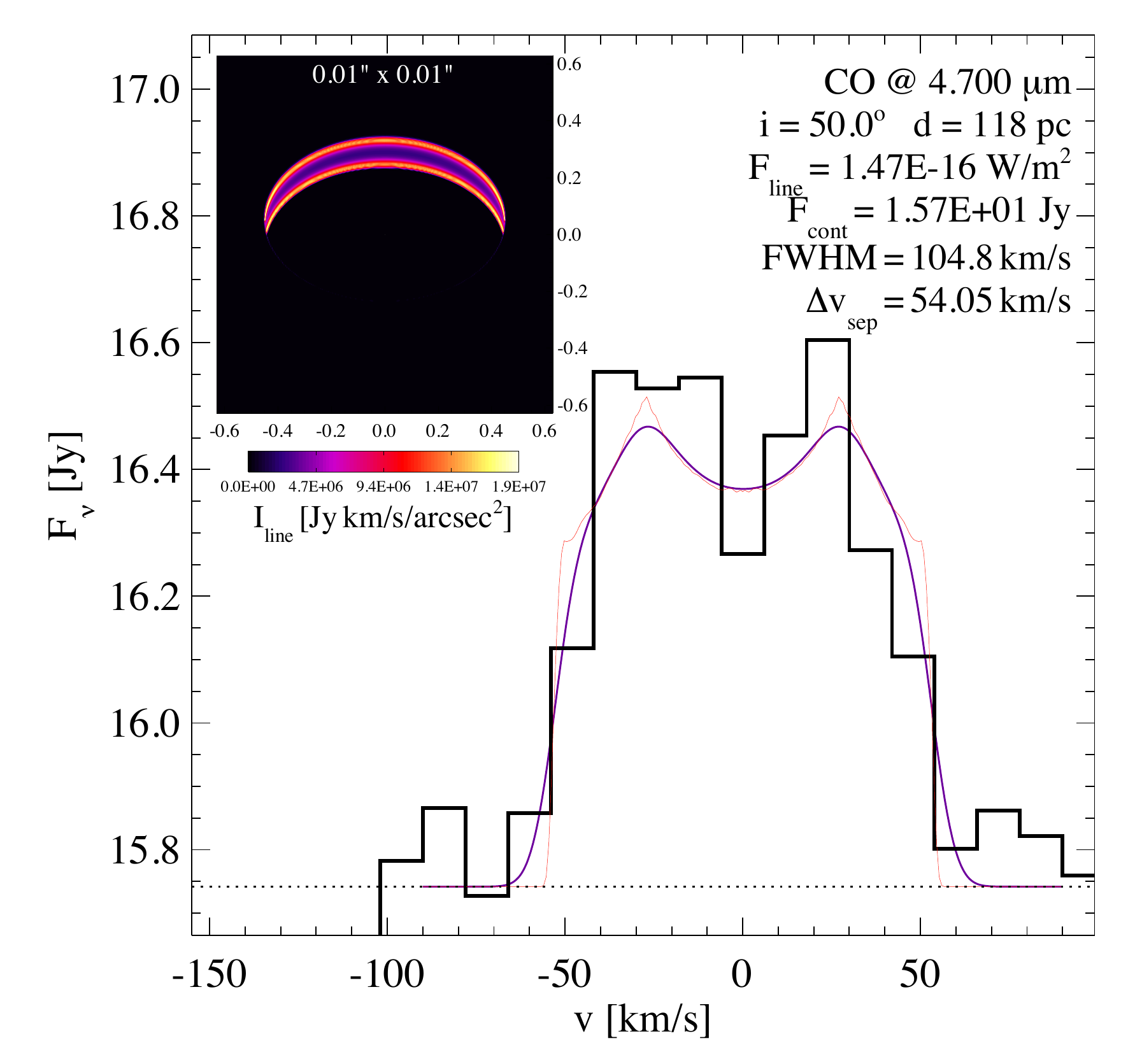}
\end{minipage}
\begin{minipage}[r]{.32\textwidth}
\includegraphics[width=\textwidth]{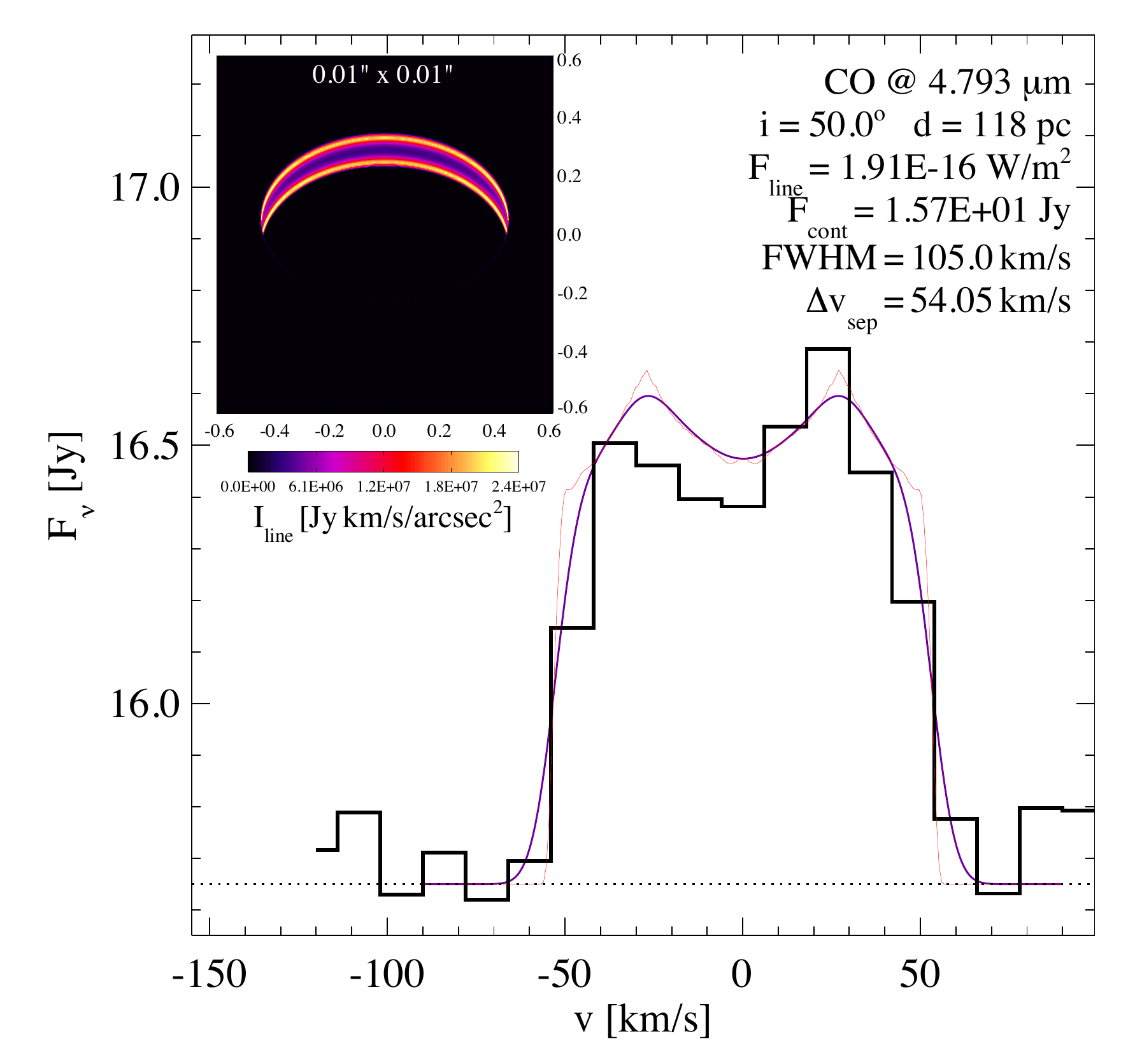}
\end{minipage}
\begin{minipage}[r]{.32\textwidth}
\includegraphics[width=\textwidth]{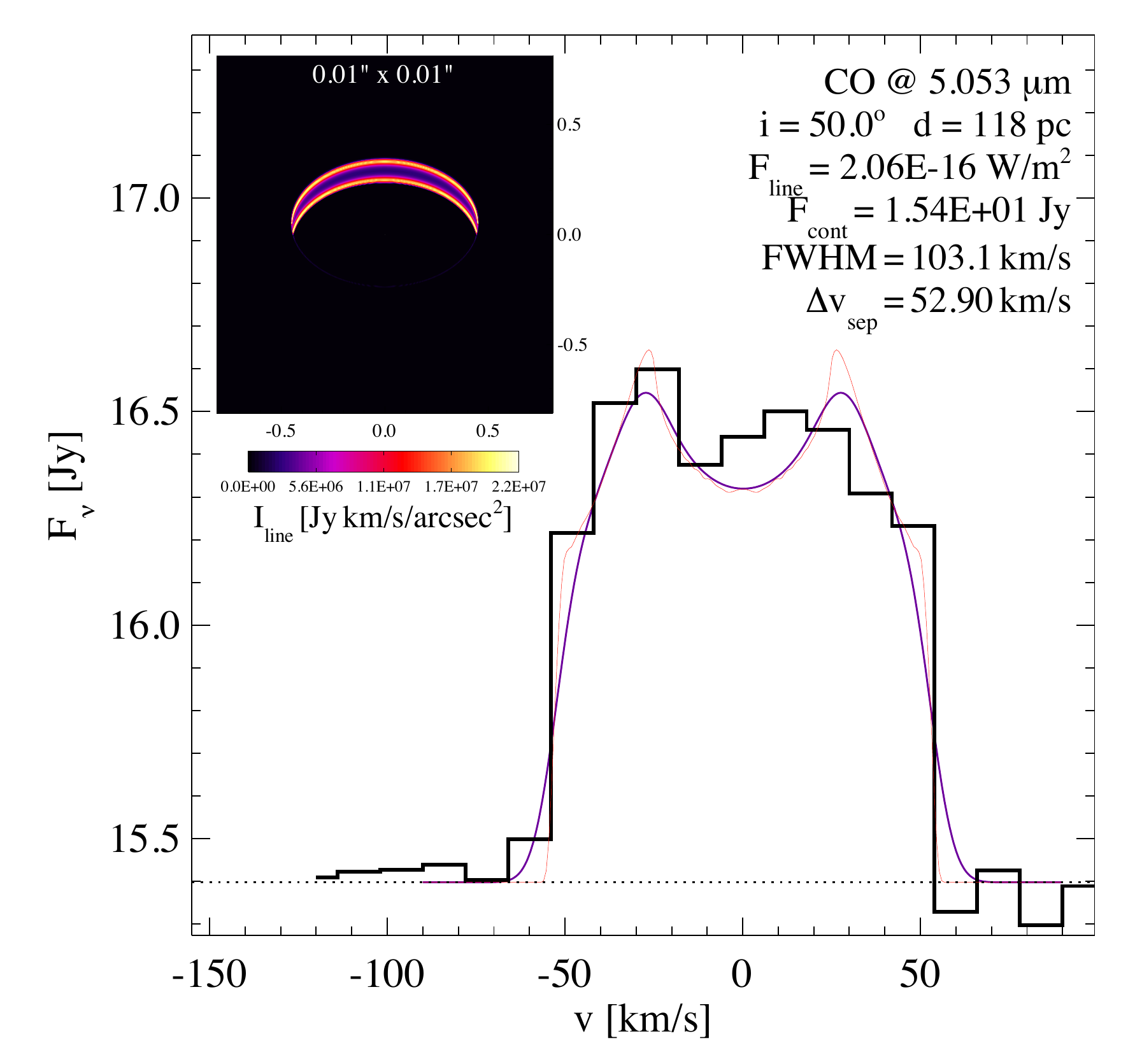}
\end{minipage}
\caption{Modelled line profiles with and without convolution and noise. From left to right: v(1--0)P04, v(1--0)P14, and v(1--0)P37. The black lines in the upper three panels show the model line profiles convolved with the CRIRES resolution (R=100000) and an applied signal to noise ratio of 100 on the continuum, while the un-convolved model line profiles are shown in red. The black lines in the lower three panels show the model line profiles convolved with the NIRSPEC resolution (R=25000) and an applied signal to noise ratio of 100 on the continuum. The line wavelength, flux, FWHM and peak separation of the (un-convolved) model are reported on the plot. }
         \label{fig:mod_line_CRIRES}
\medskip
\begin{center}$
\begin{array}{cc}
  \includegraphics[width=.327\textwidth]{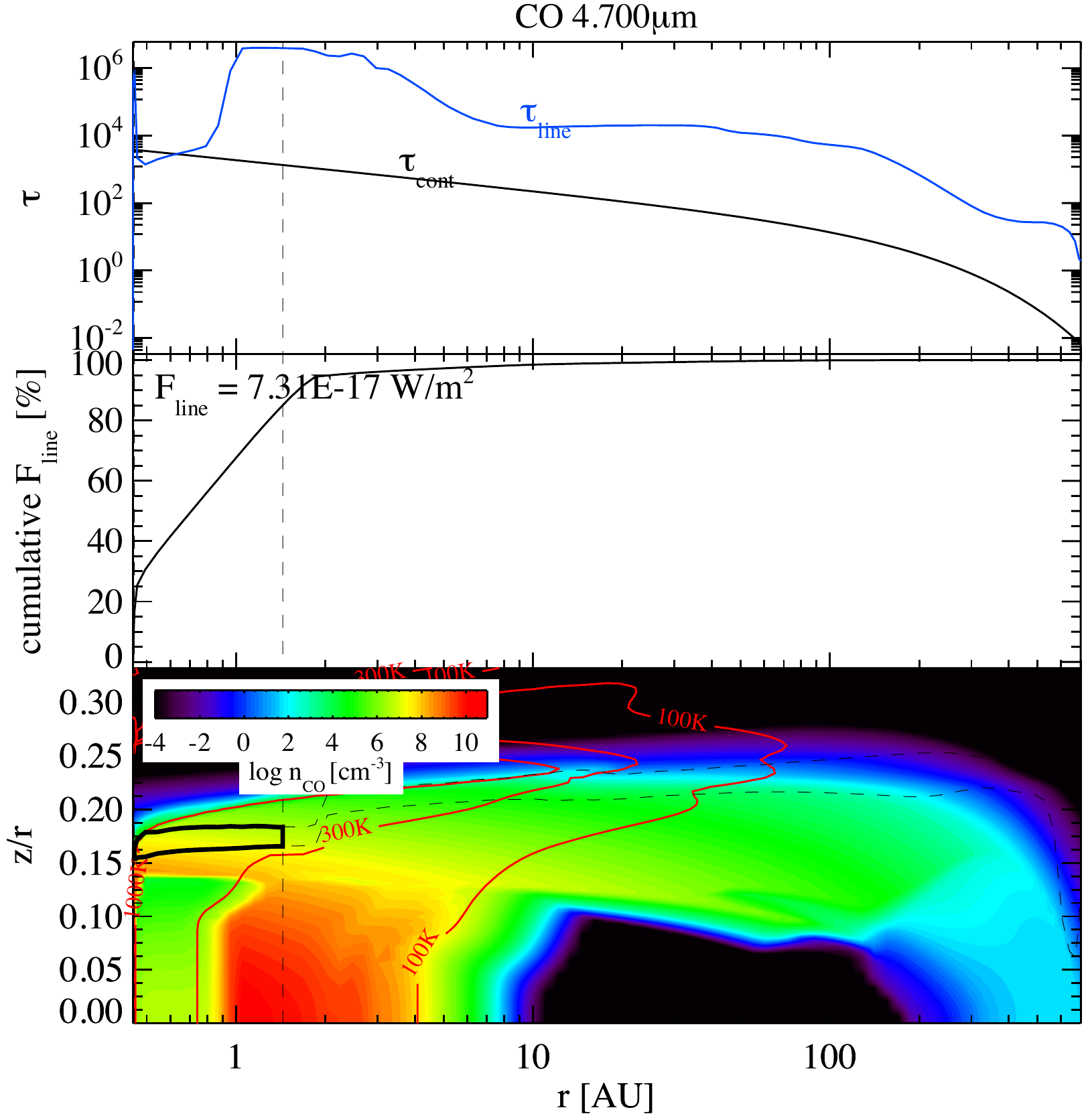}
  \includegraphics[width=.327\textwidth]{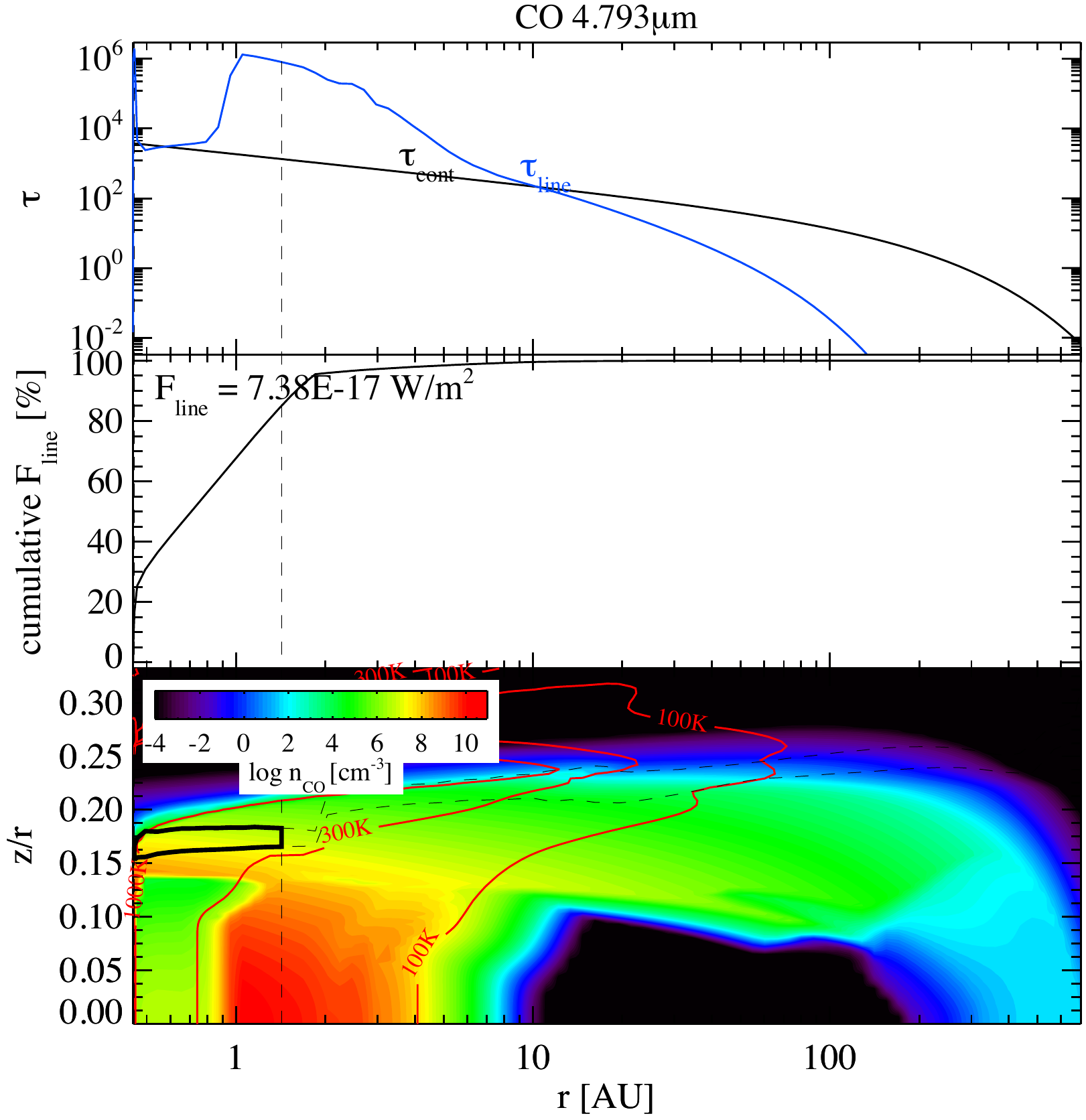}
  \includegraphics[width=.327\textwidth]{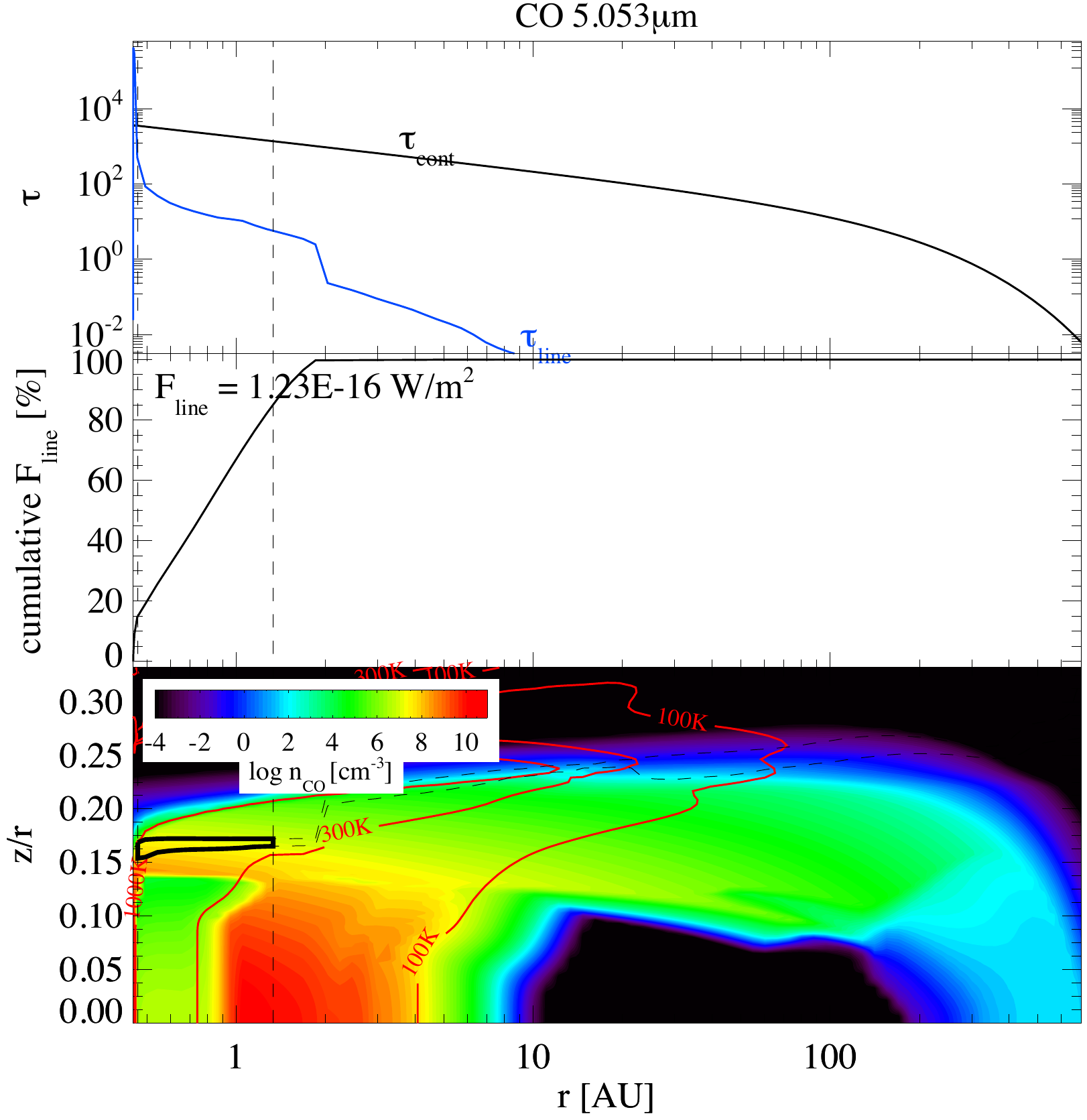}\\
 \end{array}$
\end{center}
\caption{Line and continuum optical depth, the cumulative line flux, and the CO density, as a function of radius, for the three modelled lines v(1--0)P04, v(1--0)P14, and v(1--0)P37. The line wavelengths and fluxes from simple vertical escape probability are indicated on the plot. Vertical dashed lines indicate the radii within which 15\% and 85\% of the total flux is emitted, the black box indicates the radial and vertical region from which 50\% of the flux originates, and the red contour lines represent gas temperatures of 100\,K and 300\,K.}
         \label{fig:mod_nh_c}
%\end{figure*}

\end{figure*}

\subsection{Line profile comparison} \label{sec:modres}
We derive line flux, FWHM and peak separation from the modelled CO ro-vibrational line profiles. Line shapes in the disc model do not vary strongly with $J$. The peak separations in the modelled lines are $\sim$50\,km/s, and the NIRSPEC resolution could thus be good enough to separate the peaks, depending however on the size of the depression between the peaks.
In the upper panels of Fig. \ref{fig:mod}, the peak separation, the FWHM, and the line fluxes are shown for selected line transitions (the observed line sample and a few extra lines to fill in gaps in $J$ coverage) from the model. The lower panels show these same plots with all transitions calculated in LTE. For comparison we also show the observed peak separations, FWHM, and line flux. To obtain a more realistic comparison between model and observation, we add appropriate noise (S/N$\sim$100 on the continuum) and convolve the model line profiles with the spectral resolution of the relevant instruments (CRIRES: 3\,km/s, NIRSPEC: 12\,km/s) to create simulated CRIRES and NIRSPEC line profiles from the modelled lines. The simulated line profiles made for three representative CO ro-vibrational transitions, v(1--0)\,P04, v(1--0)\,P14, and v(1--0)\,P37, are shown together with the original model line profiles in Fig. \ref{fig:mod_line_CRIRES}. 
 These three lines are chosen since they exemplify a low $J$, a mid $J$, and a high $J$ line, and since they are detected in both observational data sets, N01/02 and C12.

The simulated CRIRES and NIRSPEC line profiles from the models have FWZI around 110--120\,km/s (measured by eye), similar to the observed FWZI (The FWZI of the lines is determined by the inner radius of the disc. In the model that is set to 0.45 au, while values derived from observations lie in the range 0.2--0.55 au \citep{renard2010, eisner2009, tannirkulam2008, benisty2010}.). The FWHM values predicted by the model are a factor 1.3 too wide compared to observed lines from N01/02 (Fig. \ref{fig:mod}). 
The line fluxes predicted by the model match the observed low $J$ line fluxes (N01/02 and C12) but the high $J$ line fluxes are up to a factor four too faint and the overall shape of the line flux versus $J$ curve is different than what is observed. The LTE model captures the line flux versus $J$ curve shape much better but the line fluxes are almost a factor ten too high.

\begin{figure*}
\centering
\begin{minipage}[l]{\textwidth}
\includegraphics[width=.9\textwidth]{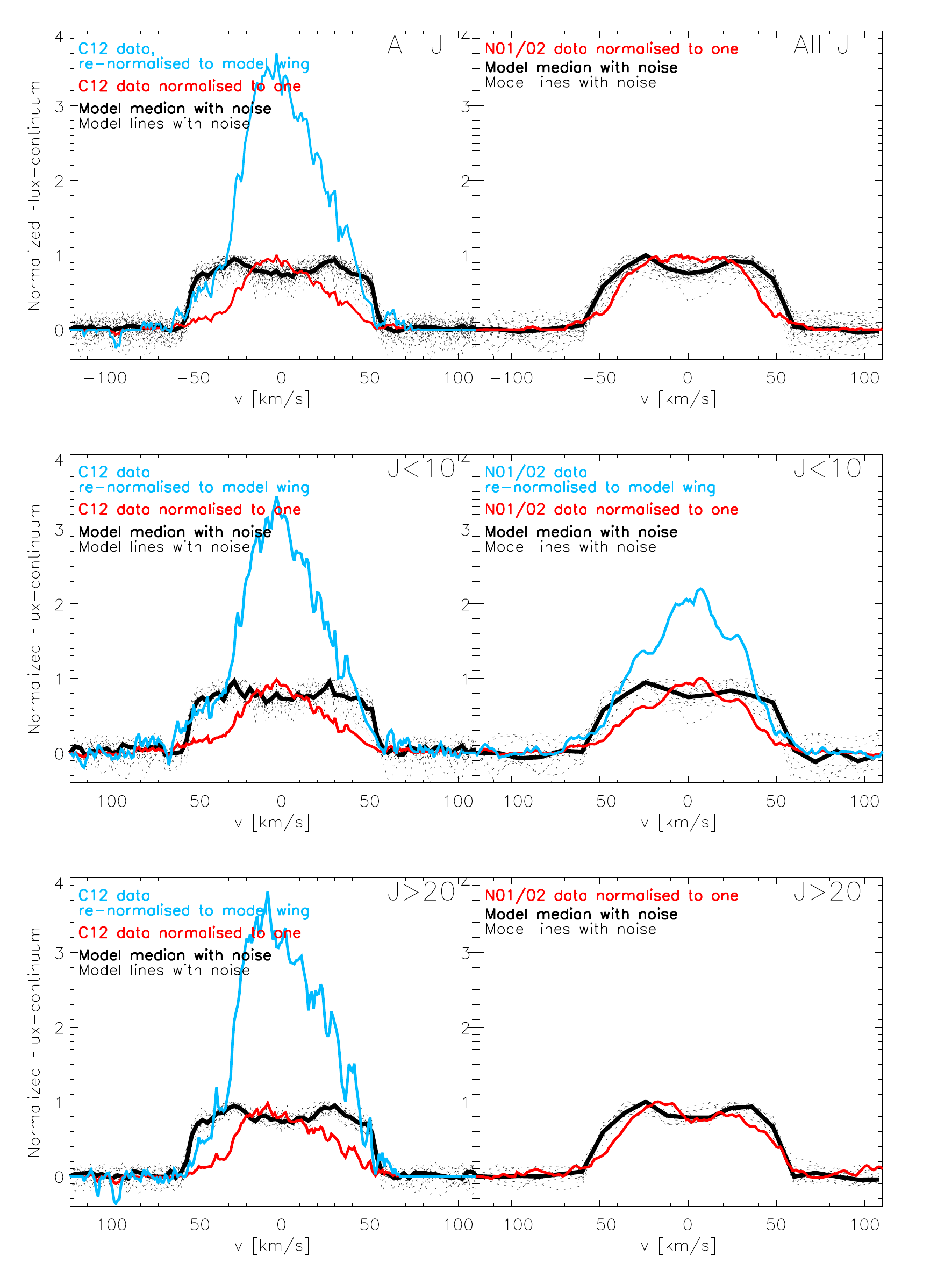}
\end{minipage}
\caption{Upper panels: Individual normalised modelled line profiles (dotted black lines) and their medians (thick black) with an applied signal to noise ratio of 100 (on the continuum) and the spectral resolution of CRIRES (left) or NIRSPEC (right). For comparison, the observed line profile median is overplotted in thick red (\textit{left}: C12, \textit{right}: N01/02). Medians created using only the low $J$ lines ($J$<10) shown in the middle panels, and medians created using only the high $J$ lines ($J$>20) shown in the lower panels.
For the C12 medians and the low $J$ N01/02 median, we additionally overplot the observed medians re-normalised instead to the line wings of the model median (thick blue) to evaluate the presence of an additional non-Keplerian component causing the central peaks. }
         \label{fig:mod_avline}
\end{figure*}

The model predicts clear double peaks for all lines (significant central depression between peaks), which is not seen for all lines in the observations. Since the CO ro-vibrational lines are not spatially resolved in observations the observed single peaks cannot be caused by emission at larger radii in the disc. Thus, the observed single peaks in both the CRIRES lines and the low $J$ NIRSPEC lines could indicate that lines from both N01/02 and C12 have a second non-Keplerian component, although much weaker in N01/02 than in C12. We do see indications of double peaks present in some observed lines (mostly high and mid $J$ lines from N01/02) and the peak separations measured from these are similar to the modelled lines.
 However, the absence of a strong double peak could also be due to many lines in the N01/02 sample suffering from low transmission at the centre of the line, due to strong telluric absorption lines (see Sect. \ref{sec:obsres}). 
The high $J$ transitions are less likely to be significantly affected by telluric absorption and lend themselves better for a comparison to the models. 
In Fig. \ref{fig:mod_avline} we show flux normalised medians from the simulated CRIRES line profiles and NIRSPEC line profiles from the model together with the corresponding observed (N01/02, C12) median line profiles. We show both medians made from all $J$ transitions and medians made from only high or low $J$ lines in the sample. We can confirm that the high $J$ median from the N01/02 sample compares well to the model. 
Looking in particular at the line wings, the median created from all lines from N01/02 and the median created from high $J$ lines from N01/02 look similar. However, comparison of the medians from C12 and the low $J$ medians from N01/02 reveals significant differences in the line wings.
In these cases, the line profiles might be multiple component profiles with a non-Keplerian, central component. Thus, the direct comparison of medians from model and from observations, both normalised to unity, can be misleading, since the model contains only a disc component. We can obtain a match between the line wings of the observed C12 lines and the line wings of the simulated CRIRES line profiles from the model by re-normalising the C12 median line profile to the wings of the model (multiplying the normalised C12, low $J$ median by 3.5, shown in blue). A similar match could also be achieved for the low $J$ median from N01/02 (multiplying the normalised N01/02 median by 2.2, shown in blue). 

The shape of a line profile is determined by how the flux builds up as a function of radius. In Fig. \ref{fig:mod_nh_c} we show a radial profile of the cumulative line flux from the model. From the peak separation of the high $J$ lines from N01/02, we infer an outer radius for the emitting region of $\sim$2.1$\pm$2.0 au, using the assumption of Keplerian rotation. 
In the model, 85\% of the emission builds up within $\sim$1.5 au. Thus, the model provides a good prediction of the emitting region.
In summary, provided there is a non-Keplerian additional component causing a central peak in the lines from C12 and in the low $J$ lines from N01/02, the model compares well to the observed CO ro-vibrational lines of HD\,163296.

\section{Discussion} \label{sec:disc}

\subsection{Observed line profile comparison}
N01/02 and C12 show clear differences in the line profile shapes that instrumental differences alone cannot explain. The single peaked nature of the C12 lines points at variability in the lines as an explanation. Furthermore, a detailed thermo-chemical disc model predicts double peaked line profiles which are only seen for high $J$ line profiles from N01/02. {A possible explanation for single peaks could be that an additional non-Keplerian CO ro-vibrational emission component is present in both N01/02 and C12, but that this component is weaker in N01/02, and hence mainly affecting low $J$ lines here.} The observed CO ro-vibrational emission is not spatially extended, hence, a Keplerian emission component from large radii in the disc cannot be invoked to explain the single peaked profiles. 

Comparing line fluxes, the overall shape or curvature in line flux versus $J$ diagram agrees well between N01/02 and C12, and the line flux values are also similar. Hence, the line fluxes alone do not indicate any kind of variability in the lines. However, the flux calibration, is rather uncertain, around 30\% for C12 and 20\% for N01/02, and thus the similarity in line fluxes does not necessarily exclude variability.

{Spectra from multiple epochs were combined to obtain the N01/02 dataset since no indication of variability between line profiles was detected beyond the noise. No independent flux calibration exists between the multiple epochs; the N01/02 data were neither obtained nor analysed with the intention to study variability.
The same is true for the C12 data. Hence, there is now a clear need for new systematically collected data to better understand the variability of this source (we discuss this further in Sect. \ref{sec:disc:var})}.

\subsection{Model comparison}
Considering the fact that this model was not set up to fit the CO ro-vibrational lines initially, this comparison provides a strong test. For high $J$ lines from N01/02, the peak separations compare well and the FWHM are near the observed values. If all other observed lines are in fact affected by the presence of an additional non-Keplerian component (that is not included in the model), the FWHM or peak separation of the remaining observed lines cannot be expected to match the Keplerian lines of the model. 

The individual line fluxes predicted by the model are similar to the observed values up to J$\sim$26. The modelled high $J$ lines fluxes deviate by up to a factor 4 from the observed ones. The overall shape of the flux versus $J$ diagram is not well fitted by the model. The modelled lines calculated in LTE give a much better shape prediction for the flux versus $J$ diagram, but the line fluxes are in this case almost a factor of ten too high,a difference that cannot be explained by flux calibration uncertainties (C12, 30\% systematic flux calibration uncertainty, N01/02, 20\% systematic flux calibration uncertainty).
{The difference in line strength between the (non-LTE) model and observations may point to a different vertical structure in the inner disk model and/or the excitation of the lines (thermal and fluorescent). However, the difference in flux versus $J$ curvature could arise from an additional non-Keplerian component that is more pronounced at low $J$ compared to high $J$ lines (also responsible for the single peak/flat topped difference between high and low $J$ lines); this is not captured in the current model.}

\subsection{Variability of the CO ro-vibrational lines} \label{sec:disc:var}
\begin{figure}
\begin{center}$
\begin{array}{cc}
  \includegraphics[width=.45\textwidth]{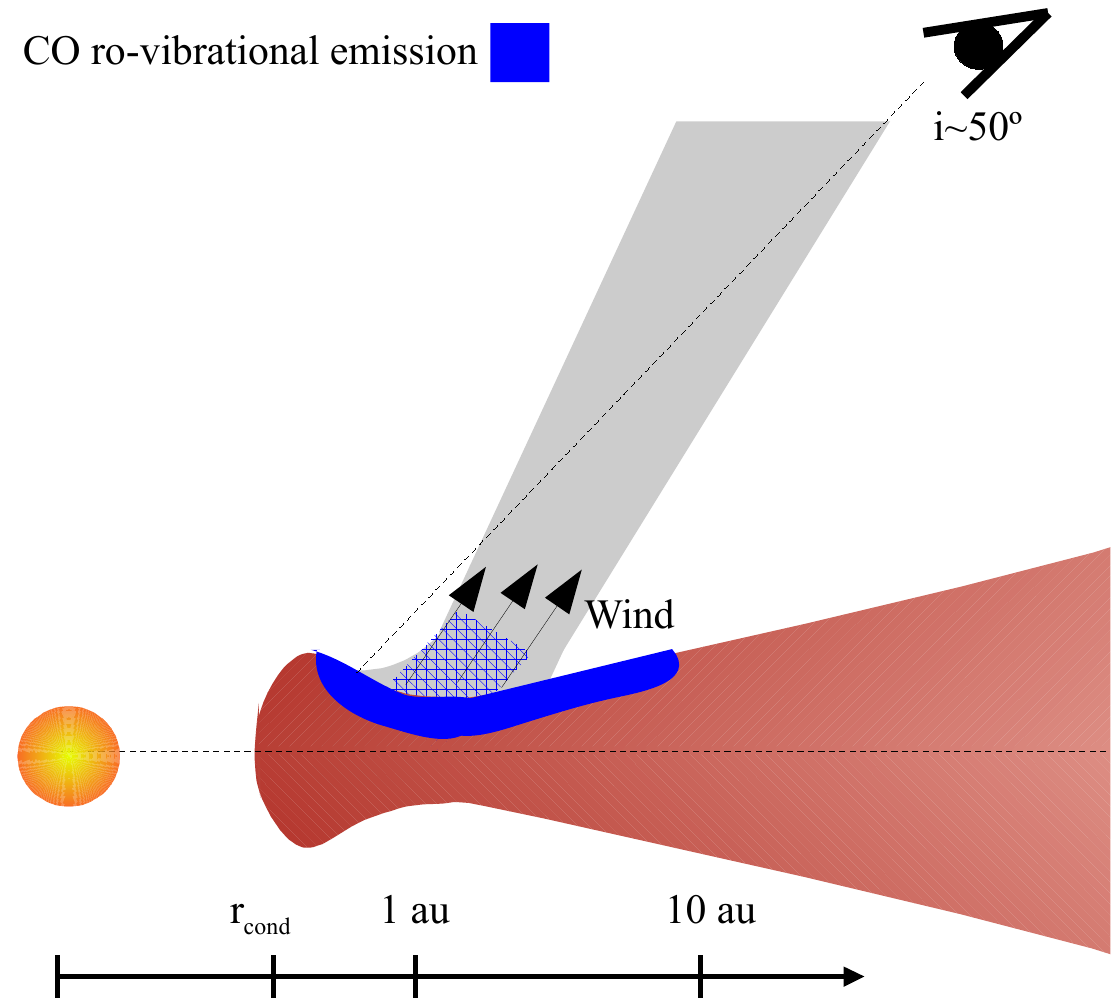}\\
\end{array}$
\end{center}
\caption{Sketch of the geometry of HD\,163296 and the disc wind and how this could relate to the variability of the CO ro-vibrational lines \citep[Inspired by][]{ellerbroek2014}. The blue areas indicate the possible CO ro-vibrational emission regions. The hatched blue area indicates CO ro-vibrational emission from the wind just above the disc.}
         \label{fig:disc_outflow_sketch}
\end{figure}

The variability of HD\,163296 is documented in the literature \citep{sitko2008,ellerbroek2014}.
\citet{ellerbroek2014} suggest a scenario where dust clouds are launched above the disc plane and cross the observers line of sight.
Klaassen et al. (2013) detected a rotating molecular disc wind from HD\,163296 using ALMA. The authors interpret this as a low velocity wind (<25\,km/s) launched at a few au. \citet{bast2011} find that broad single peaked CO ro-vib profiles observed towards T Tauri stars can be explained by the presence of a wind. Hence, this disc wind detected from HD\,163296 could be responsible for the non-Keplerian component that we suggest (velocities $\pm$30\,km/s). I.e., a low velocity molecular outflow containing CO, could emit CO ro-vibrational emission as it is launched above the usual disc emitting region, similar to what \citet{garcialopez2015} suggested for the Br$\gamma$ line. Instabilities in the inner disc could lead to episodic disc winds of varying strength resulting in e.g. stronger non-Keplerian components present in the C12 data compared to the N01/02 data. A wind component to the lines would be expected to show blue shifted profiles \citep[e.g.][]{alexander2008b,pascucci2009,pascucci2014}, consistent with the blue-shifted peak that we see in most of the velocity profiles from C12 (see Fig. \ref{fig:aalines}). Fig. \ref{fig:disc_outflow_sketch} shows a schematic view of the combined geometry of the disc, the wind and the CO ro-vibrational emission. \citet{pontoppidan2011} showed with their wind model that if the line of sight falls close to the angle of the wind column, self-absorption can occur in the line profile. The fact that no CO ro-vibrational self-absorption is detected in lines from either N01/02 or C12, puts strong constraints on the wind model. With the disc inclination of 50$\degree$ the wind column has to be launched at minimum 60$\degree$--70$\degree$ (with respect to the disc plane), almost vertically from the disc. {\citet{pontoppidan2011} showed with his wind model that if the disc wind is viewed close to face on, no self-absorption will be seen \citep[Fig. 12. in][]{pontoppidan2011}.} If the wind column is at lower inclinations, crossing the line of sight of the observer, the CO gas has to have either no sharp temperature gradient (colder gas would cause absorption), or the CO would have to dissociate before it can significantly cool.

New observations of the CO ro-vibrational lines from HD\,163296 collected with either NIRSPEC/Keck or CRIRES/VLT, that contribute additional epochs to the line profile comparisons, are necessary in order to clarify the cause for variability. New observations from one of these instruments would also allow a cleaner line profile comparison of lines from only one instrument, removing all instrumental differences leaving just the differences caused by true variability.
{Since the comparison of the four seperate datasets used in \citet{salyk2011} does not suggest any significant variations within a year, a preferred follow up programme would be a long term monitoring collecting CO ro-vibrational data from HD163296 e.g. once per year. It would furthermore be crucial to have simultaneously collected continuum data, in order to match possible CO ro-vibrational line variations to continuum variations, and better understand the physical connection between gas and dust.}
Given the difficulties with the near-IR observational window, observations of Neon fine structure lines ([Ne\,\textsc{ii}] at 12.81\,$\mu$m and [Ne\,\textsc{iii}] at 15.55\,$\mu$m) could be an alternative. \citet{Baldovin2012} detected and spectrally resolved the [Ne\,\textsc{ii}] in seven stars using VLT/VISIR, where one was the first detection of [Ne\,\textsc{ii}] from a Herbig Ae/Be star (V892 Tau).  
The [Ne\,\textsc{ii}] lines are complementary to the CO ro-vibrational lines since they probe vertical heights above the molecular layer over a range of radial distances \citep[out to 10--15\,au,][]{glassgold2007}. Hence, if a molecular outflow is causing variable non-Keplerian CO ro-vibrational emission components, the variability could also be detectable in the [Ne\,\textsc{ii}] line.

The $M$-band magnitude for HD163296 seen in the N01/02 data is close to the brightest state reported (de Winter et al. 2001), while that seen in the C12 data is $\sim$0.4 magnitudes brighter than the brightest state. In the $L$-band the amplitude of the variability does not exceed 0.2--0.3 magnitudes (Ellerbroek et al. 2014). Hence, a 0.4 magnitude increase in the $M$-band, compared to the brightest state is unlikely and it is more likely that the flux calibration of C12 is inaccurate (uncertainty closer to 50\%). With this, one could envision an alternative scenario for the line profile variability where the continuum flux from the N01/02 data is stronger than that from the C12 data. In that case the peak line flux from the two data sets might match, and the line variability would be caused by a broad wing component dominant in the N01/02 data, in particular for high and medium $J$ lines. Such an additional component could be caused by episodic accretion onto the disc (seen in the case of T Tauri stars). The enhanced accretion creates funnel flows seen as broad double peaked lines \citep{najita2003}. 

Variable accretion has in fact been noted by \citet{mendigutia2013} for HD\,163296. From their comparison with previous results they found that the accretion rate of HD\,163296 is more or less constant on timescales of days to months. However, the accretion rate derived using data from 2011--2012, was found to have increased by more than 1 dex over a timespan of 15 years. Thus, the C12 dataset was collected during a high accretion rate period, and should thereby be displaying this additional line wing component. Yet, opposite to what we would expect from the high accretion rate, the lines from the C12 data are narrower than lines from the earlier epoch, N01/02. Hence, an additional line component due to episodic accretion does not offer a meaningful explanation to the observed line variability.

\section{Conclusions} \label{sec:conc}
We compared for the first time CO ro-vibrational lines from high-resolution NIR spectra for HD\,163296, collected at different epochs (2002,2012) with different instruments (NIRSPEC/Keck, CRIRES/VLT), separated by $\sim$10 years. We find significant differences in line shapes. In particular, double peaks are only present in one epoch, and mainly in high $J$ transitions. FWHM measurements are significantly wider in the 2002 epoch. Comparison of flux calibrated lines indicates that line wings could be similar at the two epochs and an additional component at lower velocities likely causes the FWHM difference and the peak differences. However, the reliability of the flux calibration is limited (errors of $\sim$50\% for C12 and $\sim$20\% for N01/02).

We use an existing disc model \citep{Tilling2012} to derive modelled CO ro-vibrational lines for HD\,163296.
 The model provides good predictions of the peak separation of the high $J$ lines from N01/02, and good predictions for the line fluxes of the low and mid J lines (from both N01/02 and C12). For all observed lines from C12 and the lower $J$ lines from N01/02, the peak separation (predicted by the model) is not seen (single peaks or ambiguous flat tops), and the modelled FWHM are a factor 1.7 larger than in C12 and a factor 1.3 larger than in N01/02.

We propose that an additional non-Keplerian component of the CO ro-vib\-ra\-tion\-al emission is present in the C12 data set. In the epoch of the N01/02 lines, the non-Keplerian component is also present, but significantly weaker. The line flux contribution of the proposed non-Keplerian component decreases with $J$ for both data sets, and is thus only detectable for low $J$ lines in the N01/02 data set.
This would explain why the model underestimates FWHM, and peak separation, for low and mid $J$ lines.
The suggested non-Keplerian component in the observed lines, seem to be present at similar velocities in all cases ($\pm$30) and not detectable at high $J$ in N01/02.
We suggest that this non-Keplerian component could be due to the molecular disc wind detected by Klaassen et al. (2013).
Further observations are necessary to confirm the nature of the variability seen for these lines  (with CRIRES/VLT or NIRSPEC/Keck) and the presence of a variable disc wind component. The latter could be done with or VISIR/VLT observations of the [Ne\,\textsc{ii}] line.

\section*{Acknowledgements}
The authors thank S. Brittain and C. Salyk, for providing the additional CO ro-vibrational data sets used in this paper and for helpful comments and suggestions.
IK, WFT, and PW acknowledge funding from the EU FP7-2011 under Grant Agreement no. 284405. 

\bibliographystyle{mnras}
\bibliography{reference}

% Don't change these lines
\bsp	% typesetting comment
\label{lastpage}
\end{document}